\def\draftversion{false}

\RequirePackage{ifthen}
\ifthenelse{\equal{\draftversion}{true}}{
  \documentclass[aps,pra,10pt,galley,amsmath,amssymb,nofootinbib,
     superscriptaddress]{revtex4}
}{
  \documentclass[aps,pra,10pt,twocolumn,amsmath,amssymb,
     superscriptaddress,longbibliography,nofootinbib]{revtex4-1}
}

\usepackage{graphicx}
\usepackage[usenames,dvipsnames]{color} 
\usepackage{bm} 
\usepackage{soul} 
\usepackage{mathtools}

\ifthenelse{\equal{\draftversion}{true}}{
  \marginparwidth 2.7in
  \marginparsep 0.5in
  \newcounter{comm} 
  \def\commnext{\stepcounter{comm}}
  \def\commtext{{\bf\color{blue}[\arabic{comm}]}}
  \def\commmar{{\bf\color{blue}[\arabic{comm}]}}
  \def\dvm#1{\commnext\marginpar{\small DV\commmar: #1}\commtext}
  \def\scm#1{\commnext\marginpar{\small SC\commmar: #1}\commtext}
  \def\amm#1{\commnext\marginpar{\small AM\commmar: #1}\commtext}
  \def\ism#1{\commnext\marginpar{\small IS\commmar: #1}\commtext}
  \def\srm#1{\commnext\marginpar{\small SR\commmar: #1}\commtext}
  \def\mlab#1{\marginpar{\small\bf #1}}
  
  \newcommand{\seclab}[1]{\label{sec:#1}{\Red{\small\;\;[sec:~#1]}}}
  \newcommand{\eqlab}[1]{\Red{\hbox{\small\;\;[#1]}}\label{eq:#1}}
  \newcommand{\figlab}[1]{\Red{\hbox{\small\;\;[fig:~#1]}}\label{fig:#1}}
}{
  \def\dvm#1{}
  \def\scm#1{}
  \def\amm#1{}
  \def\ism#1{}
  \def\srm#1{}
  \def\mlab#1{}
  
  \newcommand{\eqlab}[1]{\label{eq:#1}}
  \newcommand{\seclab}[1]{\label{sec:#1}}
  \newcommand{\figlab}[1]{\label{fig:#1}}
}

\newcommand{\beq}{\begin{equation}}
\newcommand{\eeq}{\end{equation}}
\newcommand{\bea}{\begin{eqnarray}}
\newcommand{\eea}{\end{eqnarray}}
\newcommand{\nn}{\nonumber\\}
\newcommand{\eq}[1]{Eq.~(\ref{eq:#1})}
\newcommand{\Eq}[1]{Equation~(\ref{eq:#1})}
\newcommand{\eqs}[2]{Eqs.~(\ref{eq:#1}) and (\ref{eq:#2})}

\newcommand{\eqr}[2]{Eqs.~(\ref{eq:#1}-\ref{eq:#2})}

\newcommand{\fref}[1]{Fig.~\ref{fig:#1}}
\newcommand{\Fref}[1]{Figure~\ref{fig:#1}}

\newcommand{\sref}[1]{Sec.~\ref{sec:#1}}
\newcommand{\srefs}[2]{Secs.~\ref{sec:#1} and \ref{sec:#2}} 

\newcommand{\ket}[1]{\vert#1\rangle}

\newcommand{\ip}[2]{\langle#1\vert#2\rangle}
\newcommand{\me}[3]{\langle#1\vert#2\vert#3\rangle}
\newcommand{\ev}[1]{\langle#1\rangle}
\def\Re{\mathrm{Re}}

\def\Tr{\mathrm{Tr}}
\def\z2{$\mathbb{Z}_2$}

\def\phm{\phantom{-}}

\newcommand{\code}[1]{\textsc{#1}}

\def\k{{\bf k}}
\def\R{{\bf R}}

\def\r{{\bf r}}

\def\rrbar{\bar{\r}}

\def\a{{\bf a}}
\def\b{{\bf b}}

\def\Qc{Q_{\rm c}}

\def\bnk{_{n\k}}

\def\bmu{_{\mu}}

\def\bmn{_{mn}}

\def\btile{_{\rm tile}}
\def\bflake{_{\rm flake}}

\def\bxy{_{xy}}

\def\msk{\hspace{0.8pt}}

\def\pion{^{\msk\rm ion}}
\def\pel{^{\msk\rm el}}

\def\pT{^{\msk\rm T}}
\def\pR{^{\msk\rm R}}
\def\pTR{^{\msk\rm TR}}
\def\pI{^{\msk\rm I}}

\def\0{\mathbf{0}}

\def\half{\tfrac{1}{2}}

\def\taub{\bm{\tau}}

\def\rhobar{\bar{\rho}}

\def\barr#1{\bar{\bar{#1}}}

\def\ts{\hspace{0.7pt}}

\def\I{\textrm{I}}
\def\T{\textrm{T}}

\def\Z{\textrm{R}}
\def\pI{^{\ts\I}}
\def\pT{^{\ts\T}}

\def\pR{^{\ts\Z}}

\def\eps{\epsilon}

\def\cP{{\cal P}}
\def\cQxy{{\cal Q}_{xy}}
\def\cO{{\cal O}}

\def\xhat{\hat{\bf x}}
\def\yhat{\hat{\bf y}}
\def\zhat{\hat{\bf z}}

\def\yop{\hat{y}}
\def\ybar{\bar{y}}

\def\gg{\tilde{g}}
\def\cB{{\cal B}}

\def\tref{\mathfrak{t}}

\begin{document}


\title{Quadrupole moments, edge polarizations, and corner charges in the Wannier
representation}

\author{Shang Ren}
\affiliation{
Department of Physics \& Astronomy, Rutgers University,
Piscataway, New Jersey 08854, USA}

\author{Ivo Souza}
\affiliation{Centro de F{\'i}sica de Materiales,
  Universidad del Pa{\'i}s Vasco, 20018 San Sebasti{\'a}n,
  Spain} \affiliation{Ikerbasque Foundation, 48013 Bilbao, Spain}

\author{David Vanderbilt}
\affiliation{
Department of Physics \& Astronomy, Rutgers University,
Piscataway, New Jersey 08854, USA}


\begin{abstract}

The modern theory of polarization allows for the determination of the
macroscopic end charge of a truncated one-dimensional insulator,
modulo the charge quantum $e$, from a knowledge of bulk properties
alone.  A more subtle problem is the determination of the corner
charge of a two-dimensional insulator, modulo $e$, from a knowledge of
bulk and edge properties alone.  While previous works have tended to
focus on the quantization of corner charge in the presence of
symmetries, here we focus on the case that the only bulk symmetry is
inversion, so that the corner charge can take arbitrary values.  We
develop a Wannier-based formalism that allows the corner charge to be
predicted, modulo $e$, only from calculations on ribbon geometries of
two different orientations.  We elucidate the dependence of the
interior quadrupole and edge dipole contributions upon the gauge used
to construct the Wannier functions, finding that while these are
individually gauge-dependent, their sum is gauge-independent.  From
this we conclude that the edge polarization is not by itself a
physical observable, and that any Wannier-based method for computing
the corner charge requires the use of a common gauge throughout the
calculation.  We satisfy this constraint using two Wannier construction
procedures, one based on projection and another based on
a gauge-consistent nested Wannier construction. We validate our
theory by demonstrating the correct prediction of corner charge for
several tight-binding models.  We comment on the relations between our
approach and previous ones that have appeared in the literature.

\end{abstract}

\maketitle


\section{Introduction}
\seclab{intro}

From elementary electrostatics it is well known that the electric
polarization in an insulator, corresponding to the dipole density,
gives rise to bound charges at the surface.  However, the definition
of bulk dipole density is not obvious in the context of a quantum
treatment of the electron system, since the electron charge cloud is
not naturally decomposable into localized entities.  This problem was
solved by the modern theory of polarization, which can be formulated
in the single-particle context either in terms of Berry phases of the
Bloch functions, or in terms of dipole moments of Wannier functions
(WFs)~\cite{king-smith-prb93,resta-rmp94,vanderbilt-book18}.

Adopting the latter point of view, the polarization is defined
in terms of the dipole moment of the unit cell, taken to consist
of point ionic charges and the continuous but exponentially
localized charge clouds of the WFs attached to that cell.
Crucially, although gauge transformations of the Bloch functions
result in changes of both the shapes and charge centers of
the WFs, the vector sum of the Wannier centers
in one unit cell is gauge-invariant up to a lattice vector.
As a result, the polarization is well defined modulo a quantum
$e\R/V_{\rm cell}$, where $e$ is the quantum of charge, $\R$ is a
real-space lattice vector, and $V_{\rm cell}$ is the unit cell
volume.

Recently, several groups have explored generalizations of this theory
to the quadrupole and higher moments of the charge distributions in
insulating crystalline solids.  Benalcazar, Bernevig and
Hughes~\cite{benalcazar-sa17,benalcazar-prb17} introduced the concept
of ``topological quadrupole insulators," in which the corner charge is
quantized by symmetries, as examples of ``higher-order topological
insulators''~\cite{parameswaran-p17}. This work attracted considerable
attention. Several authors adopted a Wannier (or hybrid Wannier)
representation as a means to define the topological indices in such
higher-order topological
insulators~\cite{song-prl17,vanmiert-prb18,ezawa-prb18,
  khalaf-2019arxiv,li-prb20}. Attempts were put forward to derive a
formula for the corner charge, either when it is quantized by
symmetries~{\cite{vanmiert-prb18,schindler-prresearch19,watanabe-prb20,kooi-npjqm21}},
or in the more general case where it takes a nonquantized
value~\cite{trifunovic-prresearch20}.
It was shown that even common ionic
compounds such as NaCl may display a fractional corner
charge~\cite{watanabe-arxiv20}.
Other works~\cite{kang-prb19,wheeler-prb19} attempted to extend a
quadrupole-moment expression to the many-body case by making use of
Resta's position operator formalism~\cite{resta-prl98}, but these
approaches have proven to be
controversial~\cite{ono-prb19,watanabe-prb20}.
  
Most of these previous works have mainly been concerned with systems
whose symmetry quantizes the corner charges.  In the absence of
symmetry, however, it is unclear whether a robust definition of a bulk
quadrupole density, analogous to that of the electric polarization for
the dipole density, is possible, even at the single-particle
level~\cite{kang-prb19, wheeler-prb19,ono-prb19,watanabe-prb20}.  The
essential problem is that unlike the total dipole of the Wannier
charge distribution associated with a unit cell, the corresponding
quadrupole is not gauge-invariant.  In fact, the trace of the Wannier
quadrupole is essentially the spread functional that is minimized when
arriving at maximally localized
WFs~\cite{marzari-prb97,marzari-rmp12}; the very fact that it can be
minimized is a reflection of its gauge dependence.  It is not
surprising, then, that the off-diagonal elements of the quadrupole
tensor are also gauge-dependent, i.e., they vary according to the
exact locations and shapes of the WFs.  For this reason, the theory of
quadrupoles and higher multipoles is fundamentally different from the
theory of dipoles that underlies the modern theory of polarization.

Just as a bulk dipole density results in a bound surface charge, so a
bulk quadrupole density is expected to result in bound surface
polarizations and edge charges in 3D, or edge polarizations and corner
charges in
2D~\cite{zhou-prb15,benalcazar-prb17,benalcazar-sa17,trifunovic-prresearch20},
where it is understood that we refer to the polarization tangential to
the surface or edge.  Intuitively, a quadrupole density $\cQxy$ in a
2D sample results in bound 1D dipole densities $\cP_x=\cQxy$ at the
$+\yhat$-normal edge and $\cP_y=\cQxy$ at the $+\xhat$-normal edge.
It also results in an overall bound charge $\Qc=\cQxy$ at the corner
where these edges meet, but this $\Qc$ is not simply the sum of the
contributions expected from the edge polarizations.  Thus, such
definitions become quite subtle, even for simple classical charge
distributions~\cite{benalcazar-prb17,benalcazar-sa17,trifunovic-prresearch20}.

In fact, there are serious reasons to question whether the edge
polarization is a physical observable at all. We give two arguments
that it is not.  To do so, we focus on a large rectangular flake cut
from an insulating 2D crystal, and frame the discussion in terms of
spinless electrons.

First, recall that in the case of dipole densities, there is a robust
bulk-boundary correspondence in that the macroscopic edge charge
density is exactly given by the bulk polarization projected onto the
edge unit normal, modulo a quantum of one electron per edge unit
cell~\cite{vanderbilt-prb93}. This means that no adiabatic
periodicity-preserving perturbation at the edge, such as a
displacement of a sublattice of edge atoms, can have any effect
whatsoever on the edge charge density.  It is natural, then, to regard
the macroscopic edge charge density as a manifestation of a bulk
property.  The edge \textit{dipole density}, on the other hand, is
obviously modified by such edge-atom displacements, suggesting that it
is not a manifestation of a bulk property in the same sense.

Second, insofar as a 1D polarization $\cal P$ is well defined, we
would expect its time derivative $d\cP/dt$ to correspond to a
physically observable edge current. However, this is problematic in
the case of edge polarizations and currents.  For example, if the
insulating flake in question has been cut from a bulk that has some
nonzero orbital magnetization $M_{\rm orb}$ (as a consequence of
broken time-reversal symmetry), then there will be a persistent
counterclockwise current $I=M_{\rm orb}$ on each edge, forcing the
nonsensical conclusion that $\cal P$ increases linearly in time.  In
fact, even if the bulk material itself is time-reversal invariant, so
that its intrinsic orbital magnetization vanishes, Trifunovic, Ono,
and Watanabe~\cite{trifunovic-prb19} have shown that when such a
system is carried adiabatically around a parametric loop, this results
in a net circulation of current around the perimeter of the
sample. This would imply that the edge polarization can be changed by
an arbitrary amount by such an adiabatic cycle. These arguments
suggest that any attempt to define the change in edge polarization in
terms of an integrated current, as is done for the bulk polarization,
is bound to run into grave difficulties.

The arguments given above imply that there are serious difficulties
associated with attempts to define the bulk quadrupole density and
edge dipole density in a 2D system. By contrast, the macroscopic
corner charge is unambiguously a physical observable.  Thus, given
details of the geometric structure and the electronic Hamiltonian of
the 1D-periodic edges as well as of the 2D-periodic bulk, a robust
theory should be capable of correctly predicting the macroscopic
corner charges modulo $e$.

In this work, we show how to construct such a theory for the case of
centrosymmetric 2D insulators, based on a Wannier representation of
the electronic system at the single-particle level.
In our formulation, we first identify a bulk
unit cell, or ``tile,'' composed of a set of ionic positive point
charges and the charge distributions associated with a set of bulk
WFs.  The quadrupole density $\cQxy$ associated with this unit cell is
gauge-dependent, i.e., dependent on the exact locations and shapes of
WFs in the unit cell.  We also construct ``edge tiles'' consisting of
ions and WFs in a ``skin'' region close to the edge, and associate
surface polarizations $\cP$ to these edges. In our formulation the
edge $\cP$'s are defined independently of the bulk $\cQxy$, as they
must be since they depend upon the detailed form of the Hamiltonian at
the edge.  While the $\cP$'s are independent of a gauge change
localized at the edge, they are, like $\cQxy$, dependent on the choice
of \textit{bulk} WF gauge.  Nevertheless, we find that all gauge
dependence cancels out when the various contributions are summed, thus
allowing for a robust prediction of the corner charge.

Specifically, we work in the context of tight-binding models
of centrosymmetric 2D insulators whose bulk and edge
electronic structures are gapped.  We
solve for the ground-state electronic structure
in four configurations, namely the infinite bulk with 2D
periodic boundary conditions, 1D-periodic ribbons of finite width
in the $x$ direction, the same but finite in the $y$ direction,
and rectangular flakes with fully open boundary conditions.
We develop two formalisms for
computing the macroscopic corner
charge (mod $e$) from the bulk and ribbon calculations alone,
and demonstrate their success by direct calculation on the
rectangular flake.

In the course of preparing this manuscript, we became aware of related
work of Trifunovic~\cite{trifunovic-prresearch20}, in which similar
questions are addressed from a somewhat different point of view. While
that work considers more general unit cell shapes and corner
geometries than we do, the implementation was only presented for
the case of single-occupied-band models and for the isolated
molecular limit of the Benalcazar-Bernevig-Hughes
model~\cite{benalcazar-prb17,benalcazar-sa17}.
We occasionally comment on similarities and
differences below.

This paper is organized as follows. In \sref{tbd}, we
introduce an expression for the macroscopic corner charge in
terms of contributions from bulk, edge, and corner charge densities
based on a tiling approach. We explain how
quadrupole, dipole, and monopole contributions from bulk,
edge, and corner tiles, respectively, add up
to give the observable macroscopic corner charge.
In this formulation,
the electronic charge density associated with each tile is that
of the WFs attached to it, raising
questions about the dependence of the
bulk and edge
contributions on the gauge used to construct
these WFs.  This issue is addressed in
\sref{gauge}, where we show that the sum of bulk and edge
contributions is indeed gauge-invariant, even though the
individual contributions are not.
In \sref{methods}, we provide additional details about
our methodology.  Specifically, in \sref{macro-ave} we discuss how we
calculate the macroscopic corner charge directly from a finite
flake.  Then in \sref{ribbons} we present several
approaches to the construction of Wannier functions
for ribbon models, including a projection approach
(\sref{proj}) and approaches based on maximal localization applied
first transverse (\sref{nw-trans}) or parallel
(\sref{nw-long}) to the extended ribbon direction.
We then demonstrate in \sref{results} the limitations of a naive hybrid
Wannier implementation, and show that these are overcome using
the gauge-consistent projection method, for three
centrosymmetric tight-binding models at half filling. Specifically,
we consider a two-band model~\cite{zhou-prb15}, a related four-band model,
and the four-band model proposed in
Refs.~[\onlinecite{benalcazar-sa17}] and [\onlinecite{benalcazar-prb17}]
to discuss quantization of the corner charges.  In \sref{gcnw}, 
we present a nested maximally-localized Wannier
construction that also generates a consistent gauge, and working in
the context of the four-band model, show that this also provides
a correct prediction of the corner charge.
We discuss some possible generalizations of our approach and its
relation to the theory of orbital magnetization
in \sref{discuss}, and summarize in \sref{summary}.

\section{Preliminaries}
\seclab{tbd}

\subsection{General considerations from tiling}
\seclab{tiling}

We consider a centrosymmetric 2D crystalline material having a rectangular
unit cell with lattice vectors $\a=a\xhat$ and $\b=b\yhat$.
A finite sample, or ``flake,'' has been cut from this
material, and its charge density is assumed to be written as
the sum of $N_x\times N_y$ contributions from the
individual unit cells.  In the deep interior all these cells are identical,
but those near the edges and corners are modified by the presence of the
boundaries.

\begin{figure}
\centering\includegraphics[width=2.6in]{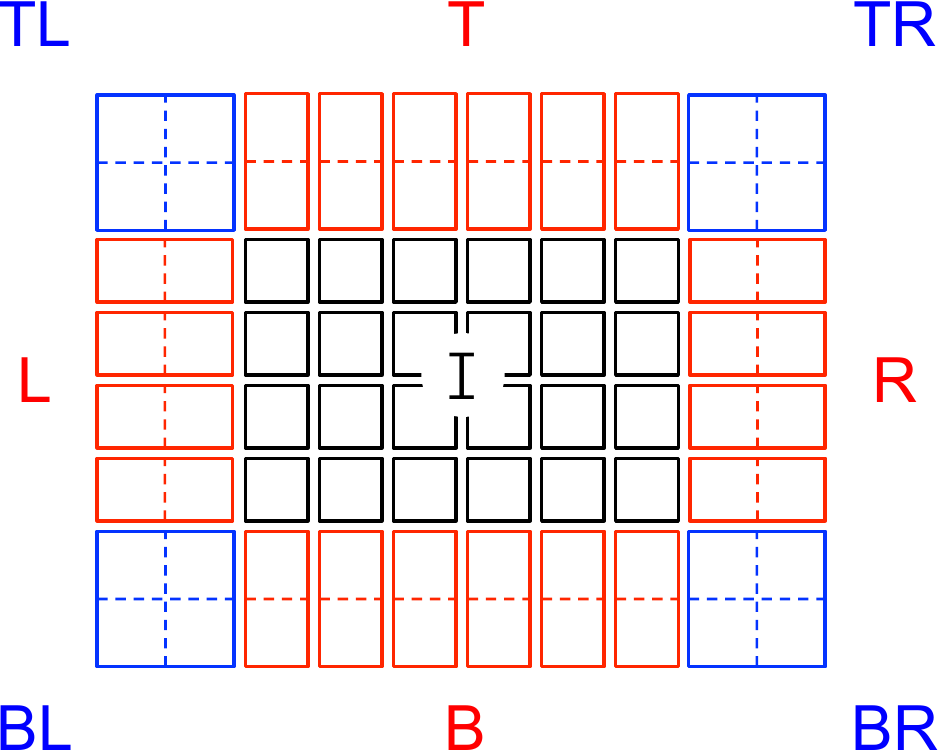}
\caption{Sketch of tiling scheme for a rectangular sample cut
  from a 2D crystal.  Small square tiles (black) correspond to
  single interior (`I') unit cells.  Rectangular edge tiles (red)
  and larger corner tiles (blue), which may extend to a depth of two
  or more cells, define the skin region. Edges are labeled as `T'
  (top), `R' (right), `B' (bottom), and `L' (left), and corners
  are labeled by combinations such as `TR' (top right).}
\figlab{skin}
\end{figure}

We identify a ``skin region'' on each edge, consisting of
$M_y$ cells at top and bottom and $M_x$ cells at left and right, where
$M_x$ and $M_y$ are chosen large enough that the deeper tiles are
bulk-like to some desired accuracy.  This is illustrated in
\fref{skin}, where $M_x=M_y=2$.
We decompose the charge density of the
flake as a whole into contributions from the interior, the four
skin regions, and the four left-over corner regions.  That is,
we write
\beq
\rho\bflake(\r) = \rho^{\rm I}(\r) + \rho^{\rm S}(\r) + \rho^{\rm C}(\r)
\eqlab{EIC}
\eeq
where the superscripts denote ``interior,'' ``skin,'' and
``corner'' contributions (black, red, and blue regions in
\fref{skin}), respectively.

The first term in \eq{EIC} is the superposition of
the identical interior tile charge densities, i.e.,
\beq
\rho^{\rm I}(\r) = \sum_{\ell_x\in I_x} \sum_{\ell_y\in I_y}
   \rho\btile\pI(\r-\ell_x\a-\ell_y\b) \,,
\eeq
where $\ell_x\in I_x$ means $M_x+1 \le \ell_x \le N_x-M_x$,
and similarly for $\ell_y\in I_y$.  The tile density $\rho\btile\pI(\r)$
represents one unit cell, but
does not have to be confined inside the rectangular boundaries of
the cell; it can leak into neighboring cells, but the
sum of these tile densities
must exactly reproduce the bulk periodic
density.  Note that $\rho\pI\btile$ is net neutral, and since we
assume inversion symmetry, we also require it to have
a vanishing dipole moment.

The second term in \eq{EIC} is a sum of four skin contributions,
$\rho^{\rm S}= \rho^{\rm T}+ \rho^{\rm R}+
\rho^{\rm B}+ \rho^{\rm L}$
(top, right, bottom, and left, respectively).  Here, for example,
the top skin contribution is
\beq
\rho^{\rm T}(\r) =  \sum_{\ell_x\in I_x} \rho\btile\pT(\r-\ell_x\a-N_y\b
) \,,
\eeq
where the ``tile'' $\rho\btile^{\rm T}(\r)$
is only one unit cell wide, but
comprises all of the $M_y$ vertically stacked cells in the top skin
region.  The density $\rho^{\rm T}\btile(\r)$
must have the property that
$\rho\pI(\r)+\rho\pT(\r)$ is identical to $\rho\bflake(\r)$
in the central region of the top edge.
Similarly, in
\beq
\rho^{\rm R}(\r) =  \sum_{\ell_y\in I_y} \rho\btile\pR(\r-\ell_y\b-N_x\a
)
\eeq
the density $\rho\btile\pR(\r)$ describes a region one cell high and
$M_x$ cells wide at the right skin region.
Since we are only interested
in neutral edges, we will require all the tiles in the skin
regions to be neutral, but they are generally not dipole-free.

The last term in \eq{EIC} is a sum of contributions from the four
corner regions,
\beq
\rho^{\rm C}(\r)= \rho\pTR\btile(\r)+ \rho^{\rm BR}\btile(\r)+
   \rho^{\rm BL}\btile(\r)+ \rho^{\rm TL}\btile(\r) \,,
\eeq
where each of these tiles is a larger one covering an entire
corner region comprised of $M_x\times M_y$ unit cells.
These corner tile densities need to make up for whatever charge
density is missing after accounting for interior and skin contributions.
For example, the top-right tile charge density is
\beq
\rho\btile\pTR(\r) = \rho\bflake(\r) - \rho\pI(\r)-\rho\pT(\r) -\rho\pR(\r)
\eeq
restricted to the vicinity of this corner.

We now focus on the top-right corner, and let $\Qc$ be the
macroscopic charge of this corner, defined as the
integral of a smoothened charge density over the corner region
(see also \sref{macro-ave}).  This is given by
\beq
\Qc=\frac{1}{ab}q\bxy\pI + \frac{1}{a}d_x\pT + \frac{1}{b} d_y\pR
+ Q\pTR
\eqlab{Qcsum}
\eeq
where
\begin{align}
q\bxy\pI &=\int x\,y\, \rho\btile\pI(\r) \, d^2r \,, \eqlab{qxy}\\
d_x\pT &=\int x\, \rho\btile\pT(\r) \, d^2r \,, \eqlab{dx}\\
d_y\pR &=\int y\, \rho\btile\pR(\r) \, d^2r \,, \eqlab{dy}\\
Q\pTR &=\int \rho\btile\pTR(\r) \, d^2r \,. \eqlab{qtr}
\end{align}
Working from right to left in \eq{Qcsum}, the contribution of $Q\pTR$
is obvious.  The contribution from the right-edge tiles is that of a
1D chain of entities of dipole moment $d_y\pR$, \eq{dy}, with density
$1/b$; this has 1D polarization $d_y\pR/b$, and thus contributes a
bound end charge of that magnitude to the top end of the chain.  The
same applies to the 1D chain of $d_x\pT$ dipoles of density $1/a$ at
the top edge via \eq{dx}.
Finally, the superposition with density $1/ab$ of identical,
neutral, dipole-free quadrupoles $q\bxy\pI$, \eq{qtr},
produces no macroscopic edge charge, but it does generate
four macroscopic corner charges: $+q\bxy\pI/ab$
at TR and BL, and $-q\bxy\pI/ab$ at TL and BR.
Combining all the contributions at the TR corner coming from
\eqr{qxy}{qtr} results in \eq{Qcsum}, which will serve as
an important basis for the remainder of this work.

\Eq{Qcsum} is claimed to hold in the thermodynamic limit,
but we expect rapid convergence with system size. The ideal
situation occurs when the tile densities all have finite support,
each vanishing outside its own local region.  In that case, the
2D periodicity relating interior tiles and the 1D periodicity
relating edge tiles guarantees that the coarse-grained charge density
$\rhobar(\r)$, obtained using the sliding window average to be
described in \sref{macro-ave}, vanishes except near the corners as
soon as $M_x$ and $M_y$ are large enough.  The corner charge
obtained by integrating $\rhobar(\r)$ over one of the corner
regions then remains unchanged by any further increase of $M_x$
or $M_y$, so that perfect convergence to the thermodynamic limit is
already achieved for modest values of $M_x$ and $M_y$.  In practice
the tile densities have exponential tails, in which case we expect
exponential convergence with sample size, an expectation that is
confirmed in the results to be presented below.

For future reference, it is useful to introduce the interior
quadrupole density
\beq
\cQxy\pI=\frac{1}{ab}\,q\bxy\pI
\eqlab{QxyI}
\eeq
and edge dipole densities
\beq
\cP_x\pT=\frac{1}{a}\,d_x\pT
\eqlab{PxT}
\eeq
for the top edge and similarly for the other three edges.  In this
language, the top-right corner charge is
\beq
\Qc=\cQxy\pI+{\cal P}_x\pT+{\cal P}_y\pR+Q\pTR \,.
\eqlab{Qcsumd}
\eeq
All quantities in \eq{Qcsumd} have units of charge $e$.

We emphasize that other definitions of edge polarizations
are possible.  First,
the definitions of the bulk quadrupole density and surface dipole
densities may differ from one formulation to another, and even
within our approach, where it can depend on the choice of tile.
Second, we would also be free to define
\beq
\begin{rcases}
\bar{\cP}_x \pT=\cP_x\pT+\half\cQxy\pI \quad\\
\bar{\cP}_y\pR=\cP_y\pR+\half\cQxy\pI \quad
\end{rcases}
\quad
\Qc=\bar{\cP}_x\pT+\bar{\cP}_y\pR
\eqlab{bar-def}
\eeq
or
\beq
\begin{rcases}
\barr{\cP}_x\pT=\cP_x\pT+\cQxy\pI \quad\\
\barr{\cP}_y\pR=\cP_y\pR+\cQxy\pI \quad
\end{rcases}
\quad
\Qc=\barr{\cP}_x\pT+\barr{\cP}_y\pR - \cQxy\pI
\eqlab{barbar-def}
\eeq
(written here for $Q\pTR=0$) in the spirit of some previous
works~\cite{benalcazar-sa17,benalcazar-prb17,trifunovic-prresearch20}.
Because we have concluded that the edge polarization is not a physical
observable, we do not think that any one of these definitions is
``more correct'' than another.\footnote{ Note, however, that the
  formulation of \eq{bar-def} has the advantage of being be easily
  generalized to treat corners subtending angles other than
  90$^\circ$, as shown in
  Ref.~[\onlinecite{trifunovic-prresearch20}].}
The reader is encouraged to
beware of different definitions of these quantities when comparing
papers from the literature.

\subsection{System of quantized charges}
\seclab{quantized}

We now assume that the charge density of the
crystal is composed of quantized charges in multiples of $e$.
This could be the fictitious world of integer point ``ions'' and
integer point ``electrons,'' but we will focus below on the case
that the electrons are represented by WFs,
each carrying charge $-e$ and exponentially localized in the
vicinity of its WF center.  The bulk tile $\rho\btile\pI(\r)$
is then constructed by choosing a set of representative ions and
WFs to include in the home cell.

The dipole moment of this interior tile is
\beq
d\bmu\pI=\int r_\mu \rho\btile\pI(\r) d^2r \,.
\eqlab{dmuI}
\eeq
Because we assumed inversion symmetry, the formal polarization,
expressed in reduced units $p_x=d_x\pI/ae$, $p_y=d_y\pI/be$, must map
to itself, modulo integers, under inversion.  There are four possible
cases in which $(p_x,p_y)$ is either $(0,0)$, $(0,\half)$,
$(\half,0)$, or $(\half,\half)$, modulo integers.  Only the first is
fully nonpolar.  The other three cases are somewhat trickier to
handle, and for these we adopt a split-basis
convention~\cite{vanderbilt-prb93}. That is, we split one or more
ions into several equal pieces, assigning these to unit cells in such
a way that the home cell is dipole-free.  For example, suppose there
is one $+e$ ion at $(0,0)$ and one Wannier center at $(a/2,b/2)$,
which would give ${\bf p}=(-\half,-\half)$. In this case we could
choose the home tile to consist of the WF density plus point ions of
charge $+e/4$ at $(0,0)$, $(a,0)$, $(0,b)$, and $(a,b)$, making for a
dipole-free home cell.  In this way, we will always arrange for
$\rho\btile\pI$ to have zero dipole moment as well as zero net charge.

We also want to restrict ourselves to neutral edges, since
otherwise the definition of a corner charge is problematic.
For the (0,0) case the edges are naturally neutral, and the
edge tile, say at the top, just consists of some overall-neutral
left-over set of ions and WFs.
For the other
cases, some edges are not naturally neutral, but they can always
be made so by a period-doubling (or, for three-fold symmetries,
period-tripling) edge reconstruction. We shall require that this
has always been done.  Since the (possibly split-basis-containing)
bulk tiles are dipole-free by construction, the (possibly enlarged)
edge tiles may also contain some fractional ionic charges,
but they will always be neutral overall.

We note in passing that a similar split-basis approach was
recently used to derive formulas for the quadrupole moment and
corner charge~\cite{watanabe-prb20}.  The authors
pointed out the gauge dependence of the quadrupole moment, but
observed that it can be removed when the system has a $C_n$
rotational symmetry ($n=3,4,6$). Mapping to a picture in which
electrons are represented by point charges located at Wannier centers,
they construct a charge-neutral and polarization-free basis by
an appropriate assignment of Wannier centers to
Wyckoff positions, an approach that is quite similar in spirit to our
tiling decomposition. The method was implemented for a variety
of model geometries in subsequent work~\cite{watanabe-arxiv20}.
However, these papers did not address the nonquantized corner charge
that can appear when the $C_n$ symmetries are absent.

\subsection{Wannier representation and choice of home cell}
\seclab{home-cell}

We now explicitly require that our 2D insulator must have a vanishing
Chern number, since otherwise the presence of gapless edge
channels would give rise to metallic boundaries, and there would be
a topological obstruction
to the construction of bulk WFs spanning the occupied bands.

Regarding the ionic charges, let the $i$'th ion in the home cell
${\bf R}={\bf 0}$ be located at $\taub_i$ and carry charge $Z_ie$.
Each ionic site
$\taub_i$ either sits on one of the four inversion centers in the unit
cell, or they appear in pairs symmetrically arranged around an
inversion center.

As for the electrons,
we assume that a smooth and periodic bulk gauge has been chosen for
the wave functions $\ket{\psi\bnk}$ of the
$n=\{1,...,J\}$ occupied bands, and that this gauge also respects
the inversion symmetry. The WFs constructed from these bands have centers
\beq
\rrbar_{\R n}=\me{\R n}{\r}{\R n}=\R+\rrbar_n\,.
\eqlab{wc}
\eeq
Since the gauge respects inversion symmetry, the $\rrbar_n$ are also
located on inversion centers or are symmetrically disposed about them
in pairs.  When we consider our flake, we assume that the
WFs of the flake become identical to these bulk WFs deep in the
interior of the flake, so that the home-cell charge distribution
$\rho\btile\pI$ is just built from these ions and WFs.  As discussed
in the previous section, this tile will always be dipole-free, even if
it requires splitting some ionic charges.

It may be useful to introduce a set of \textit{reference WF center
positions} as follows.  For each WF $\ket{\0 n}$ that sits on one
of the inversion centers, we define $\bm{\tref}_n$ to be the location
of that inversion center (i.e., equal to $\rrbar_n$); and for
every pair of WF centers symmetrically disposed about one of the
inversion centers, we again assign $\bm{\tref}_n$ 
for each of them to
be at that inversion center.  Then the interior tile charge density
\begin{align}
\rho\btile\pI(\r) = e \sum_i Z_i \delta^2(\r-\bm{\tau}_i)
    -e\sum_n |\ip{\r}{\0 n}|^2 
\end{align}
can be written as
\beq
\rho\btile\pI(\r)=\rho\btile\pion(\r)+\rho\btile\pel(\r)
\eeq
where
\beq
\rho\btile\pion(\r)=e \sum_i Z_i \delta^2(\r-\bm{\tau}_i)
   -e \sum_n  \delta^2(\r-{\bm{\tref}}_n)
\eqlab{rhoion}
\eeq
and
\beq
\rho\btile\pel(\r)=
  -e \sum_n \Big[ |\ip{\r}{\0 n}|^2-\delta^2(\r-{\bm{\tref}}_n) \Big] \,.
\eqlab{rhoel}
\eeq
The advantage of this formulation is that $\rho\btile\pion$ is a
purely classical point charge distribution that is gauge-independent,%
\footnote{To be clear, there are ``large'' or ``radical'' gauge
  transformations that shift one or more WF centers by a lattice
  vector, and ``small'' or ``progressive'' ones that can be smoothly
  connected to the identity gauge transformation.  We assume that
  the former are built into the definition of the contents of the unit
  cell, so at this point when we speak of gauge transformations, we
  mean progressive ones only.}
while all of the electronic gauge dependence is carried by $\rho\btile\pel$.

\subsection{Wannier quadrupoles and dipoles}
\seclab{quad-dip}

We are now ready to put it all together.  The ingredients
needed to compute the upper-right corner charge of \eq{Qcsum}
are given as follows.  The bulk quadrupole is
\begin{align}
q\bxy\pI&= e \sum_i\pI Z_i \tau_{ix} \tau_{iy}
         -e\sum_n\pI \me{\0 n}{xy}{\0 n} \nn
        &=q\bxy\pion+q\bxy\pel \,,
\eqlab{qxyI}
\end{align}
where the sums are over the contents of the interior (I) tile,
and $q\bxy\pion$ and $q\bxy\pel$ are the quadrupoles of the
distributions in \eqr{rhoion}{rhoel}, i.e.,
\begin{align}
q\bxy\pion &= e \sum_i\pI Z_i \tau_{ix} \tau_{iy}
  -e \sum_n\pI \tref_{nx} \tref_{ny} \,, \eqlab{qxyif}\\
q\bxy\pel &= -e \sum_n\pI \Big[ \me{\0 n}{xy}{\0 n} 
  - \tref_{nx} \tref_{ny} \Big] \,. \eqlab{qxyef}
\end{align}
The $x$ dipole of a top-edge tile is
\beq
d_x\pT = e \sum_i\pT Z_i \tau_{ix}
  -e \sum_n\pT \me{0n}{x}{0n} \,, \eqlab{dxf}
\eeq
where this time the sum is over the contents of the top edge
tile, and $\ket{l_xn}$ denotes a WF belonging to
the $l_x$$^{\rm th}$ tile along the edge.  Similarly,
\beq
d_y\pR = e \sum_i\pR Z_i \tau_{iy} 
  -e \sum_n\pR \me{0n}{y}{0n} \,, \eqlab{dyf}
\eeq
where the ket notation is $\ket{l_yn}$. Finally,
\beq
Q\pTR=e \sum_i\pTR Z_i -e N\pTR  \eqlab{Qf}
\eeq
where $N\pTR$ is the number of WFs associated with the top-right
corner tile.
Inserting \eqr{qxyI}{Qf} into \eq{Qcsum} yields the
desired expression for the top-right corner charge.

If we are only interested in the corner charge mod $e$,
 then no electronic solution
is needed for the TR region; $Q\pTR$ vanishes mod $e$
if fractional ionic charges $Z_i$ are absent, and are easily
determined if they are present. Thus, $\Qc$ can be determined
mod $e$ using only calculations on two infinite ribbons and a knowledge of
the ionic arrangement at the corner.  If we want to know $\Qc$
fully, not just mod $e$, then we also need enough information
about the electronic structure of the flake to decide the number
$N\pTR$ of occupied WFs in the corner tile.

\section{Gauge dependence of interior quadrupoles and edge dipoles}
\seclab{gauge}

In \srefs{home-cell}{quad-dip} we assumed some definite choice
of WFs providing a representation of the occupied
electronic states of the flake. Specifically, the set of all
bulk, skin, and corner WFs must be orthonormal and must exactly
span the occupied band subspace of the flake. We refer to any
particular choice of WFs as a ``choice of gauge.''  This
choice is not unique, so it is important to discuss the
gauge dependence of quantities such as $q\pel\bxy$ and
$d\pT_x$ of \eqr{qxyef}{dxf}.

A general gauge transformation corresponds to a unitary mixing
of the WFs according to
\beq
\ket{\R_1n_1}_{\rm new}=\sum_{\R_2n_2} U_{\R_2n_2,\R_1n_1} \ket{\R_2n_2}\,,
\eeq
where $U$ is unitary.  For our purposes, it is sufficient to consider
the transformation properties under infinitesimal unitary
transformations, since finite gauge transformations can always
be built up by using these as generators.%
\footnote{Strictly speaking, this only applies to ``small'' or
  ``progressive'' gauge transformations, i.e., those that can be
  continuously deformed to the identity.  ``Large'' or ``radical''
  gauge transformations that shift some WFs into a
  neighboring cell are also possible, but these would correspond
  to a different choice of tiling.}
The general form of an infinitesimal unitary operator is
$U=e^A=1+A$ for infinitesimal antihermitian $A$.  In the
bulk part of the flake, we want the WFs to retain the property
of being periodic images of each other, so we require that $A$ be
lattice-periodic, i.e., $A_{\R_1n_1,\R_2n_2}=
A_{\R_1+\R',n_1,\R_2+\R',n_2}$.  We further specialize to the
case that $A$ specifies a mixing of amplitude $\eps$ between
WF $n_1=m$ in cell $\R_1=\R$ and WF $n_2=n$ in cell $\R_2=\R+\R'$,
since more general gauge transformations can again be built up
from elementary ones such as this.

\begin{figure}
\centering\includegraphics[width=3.4in]{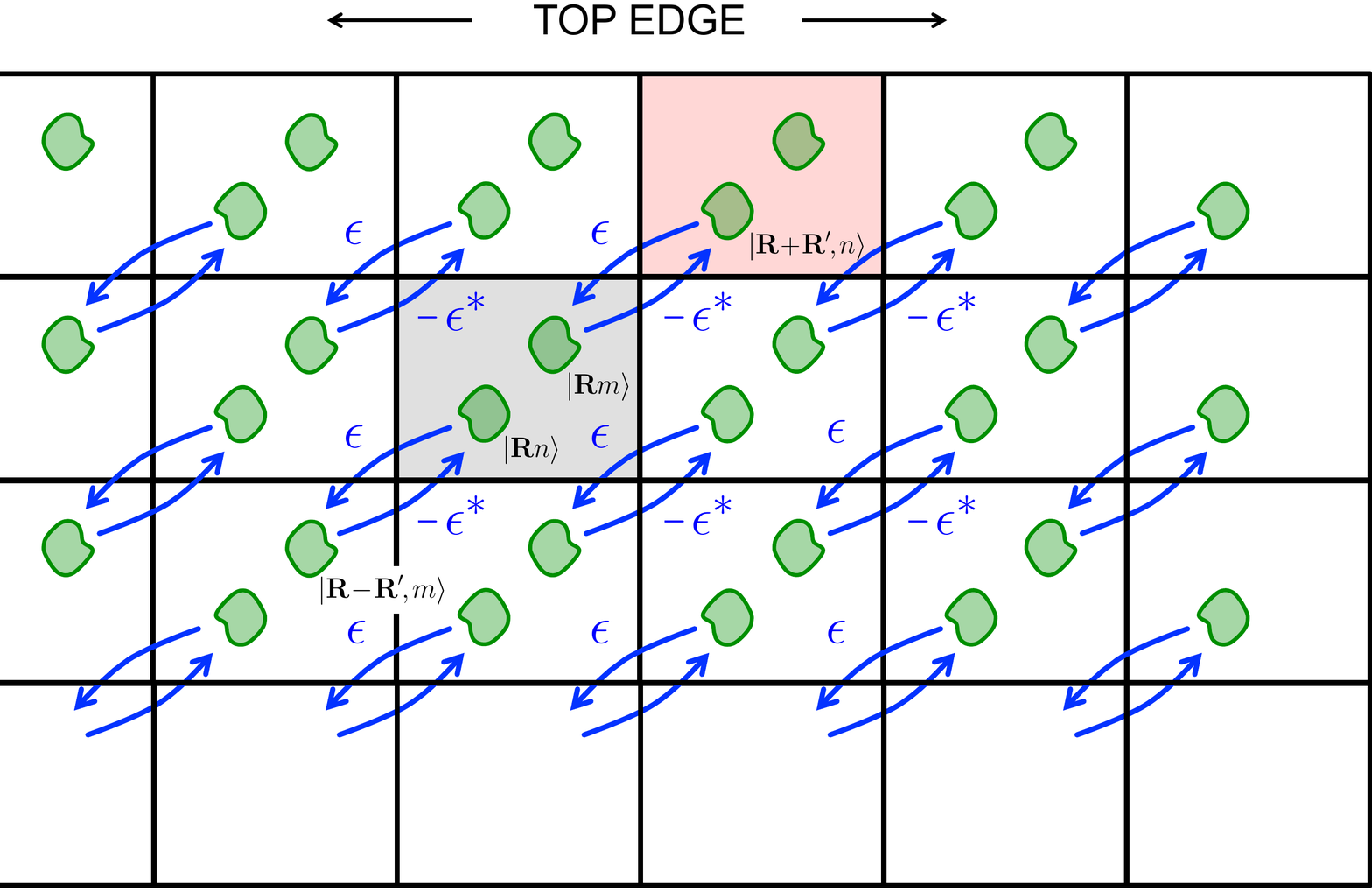}
\caption{Sketch of top edge of sample, showing mixing of
Wannier functions under the infinitesimal gauge transformation
of \eq{gc-eps}.
Gray cell is an interior cell whose dipole moment is unchanged
due to cancellation of the $\eps$ contribution from above and
the $-\eps^*$ one from below;
pink cell is a top skin cell whose dipole does shift as a
result of the unbalanced $-\eps^*$ contribution from below.}
\figlab{gt}
\end{figure}

The first-order changes in the WFs in tile $\R$ are then given by
\begin{align}
\delta \ket{\R m} &=  \eps   \ket{\R+\R',n} \,, \nn
\delta \ket{\R n} &= -\eps^* \ket{\R-\R',m} \,,
\eqlab{gc-eps}
\end{align}
with other WFs in the cell being unaffected.
The mixing pattern is illustrated in \fref{gt}.
For an arbitrary single-particle operator $\cO$,
the change of its trace over the WFs in cell $\R$ is given by
\begin{align}
\delta
\ev{\cO}\btile &= \delta \me{\R m}{\cO}{\R m}
                 + \delta \me{\R n}{\cO}{\R n} \nn
  &= 2\Re\big[ \eps\me{\R m}{\cO}{\R+\R',n} \nn
  &\hspace{1.6cm} -\eps\me{\R-\R',m}{\cO}{\R n} \big] \nn
  &= 2\Re\big[ \eps \me{\R m}{[\cO,T_{\R'}]}{\R n} \big] \,,
\eqlab{Otile}
\end{align}
where $T_{\R}$ is the operator that translates by lattice
vector $\R$.  For a lattice-periodic operator such as the
bulk Hamiltonian, the commutator in \eq{Otile} vanishes,
and the density of $\cO$ per unit cell is
 gauge-invariant.

However, we are interested in dipoles and quadrupoles, and for
these cases we have that
$[x,T_{\R}]=R_xT_{\R}$,
$[y,T_{\R}]=R_yT_{\R}$, and
$[xy,T_{\R}]=(yR_x+xR_y+R_xR_y)T_{\R}$. Using the orthogonality
of the WFs, $\ip{\R m}{\R+\R',n}=\delta_{\0\R'}\delta_{mn}$,
it follows that
\begin{align}
\delta \ev{x}\btile &=
\delta \ev{y}\btile = 0 \,, \eqlab{xandy} \\
\delta \ev{xy}\btile &= 2R'_x\Re[ \eps Y_{mn}^{\R'}]
      + 2R'_y\Re[\eps X_{mn}^{\R'} ] \,,
  \eqlab{xycell}
\end{align}
where
\begin{align}
X_{mn}^{\R}&=\me{\0 m}{x}{\R n} \,, \nn
Y_{mn}^{\R}&=\me{\0 m}{y}{\R n} \,.
\eqlab{XYdef}
\end{align}
\Eq{xandy} confirms that the dipole moment of the Wannier charge
distribution in a bulk tile is gauge-invariant, as expected
since it corresponds to the electric polarization.
Another way to see this is to compute the shifts of the
Wannier centers $\bar{x}_{\R m} = \me{\R m}{x}{\R m}=
R_x+\bar{x}_m$;
using the same methods, we obtain
\beq
\delta \bar{x}_m = - \delta \bar{x}_n = 2\Re\big[\eps X_{mn}^{\R'}\big]
\,,
\eqlab{shift}
\eeq
and similarly for $\delta \bar{y}$.  The two WF centers thus
shift by equal distances but in opposite directions, preserving
the overall cell dipole.

However, the gauge invariance of the dipole does not extend to the
quadrupole. From \eq{QxyI}, (\ref{eq:qxyI}), and (\ref{eq:xycell})
we obtain
\beq
\delta \cQxy\pI = -\frac{2e}{ab} \left(
  R'_x\Re[\eps Y_{mn}^{\R'}] +
  R'_y\Re[\eps X_{mn}^{\R'}] \right) \,.
\eqlab{Qvar}
\eeq
This shows that the bulk quadrupole moment of an interior tile
is not a gauge-invariant quantity.  In particular, this suggests
that it is not a physical observable.

Now let us concentrate our attention on the skin region,
specifically at the top edge of the flake.
The quadrupoles in this region are of no interest, since the
area of the skin region becomes negligible in the limit of a
large flake. A gauge change that is restricted only to the skin
region cannot change the dipole moment of an edge tile, by
an argument similar to that leading to \eq{xandy}.

Surprisingly, though, the dipole of an edge tile can be modified
by an \textit{interior} gauge transformation.  To see this, we
return to \fref{gt} and discuss it in the context of \eq{shift}.
Note that \fref{gt} is drawn for the case that $\R'=\ell_x\a+\ell_y\b$
with $\ell_x=\ell_y=1$,
and for simplicity we assume that the skin tile is only
one unit cell thick.  In this case, each skin tile ``donates''
a contribution $-2e\Re[\eps X\bmn^{\R'}]$ to one of the top-most
interior tiles below it, as illustrated by the blue arrow marked
$\eps^*$ pointing from $\ket{\R+\R',n}$ in the pink skin cell
to $\ket{\R m}$ in the gray interior cell in \fref{gt}.  As a
result, the shift of $\rrbar_{\R+\R',n}$ adds to the dipole of
the pink edge tile by $2e\Re[\eps X\bmn^{\R'}]$, and the shift
of $\rrbar_{\R m}$ in the gray tile makes an equal and opposite
contribution to the gray-tile dipole.  However, there is no
net change of the gray-tile dipole, since it receives a
compensating donation marked by the $-\eps^*$ arrow from
the deeper tile below it.  By contrast, no such cancellation
occurs for the pink tile, so there is a net change of its
dipole, and a resulting change by $(2e/a)\Re[\eps X\bmn^{\R'}]$
of the edge polarization $\cP_x\pT$.

This result depends crucially on the choice of $\ell_y=1$, as
in \fref{gt}, for the relative lattice vector $\R'$ involved in the
unitary mixing.  If $\ell_y=2$, then there are two uncompensated
contributions to the edge tile instead of one, and if $\ell_y=-1$,
then the transfer of dipole moment goes in the reverse direction.
Overall, then, we find that
$\delta d_x\pT=2e\ell_y\Re[\eps X\bmn^{\R'}]$, and using
\eqs{PxT}{dxf} together with $\ell_y=R'_y/b$, and applying
similar considerations to the right edge, we find that
the bulk-gauge-induced changes to the edge dipole densities are
\begin{align}
\delta \cP_x\pT &= \frac{2e}{ab}R'_y\Re[\eps X\bmn^{\R'}] \eqlab{Pxvar} \,,\\
\delta \cP_y\pR &= \frac{2e}{ab}R'_x\Re[\eps Y\bmn^{\R'}] \eqlab{Pyvar} \,.
\end{align}

Finally, as for the top-right corner tile, neither its quadrupole
nor its dipole can contribute to the macroscopic corner charge.
Moreover, its net charge density, given by \eq{Qf}, is
obviously gauge-invariant, so that $\delta Q\pTR=0$. 

Combining these contributions to \eq{Qcsumd}, we find that
the contributions from \eqs{Pxvar}{Pyvar} exactly cancel
the one from \eq{Qvar}, so that
\beq
\delta \Qc
  = \delta \cQxy\pI + \delta \cP_x\pT +\delta \cP_y\pR = 0 \,.
\eqlab{invar}
\eeq
In other words, the bulk quadrupole density and edge dipole densities
are individually gauge-dependent, but their sum is gauge-invariant
and describes a physical observable, the corner charge.
This is a major result of our work.

A crucial consequence of this result is that the corner
charge $\Qc$ can be obtained modulo $e$ from independent calculations of
$\cQxy\pI$, $\cP_x\pT$, and $\cP_y\pR$, but \textit{only} if all
three contributions are computed using the same bulk
gauge.  For example, by studying ribbons that are finite in $y$
and infinite along $x$, we can compute $\cQxy\pI$ from the charge
density of a deep interior tile, and $\cP_x\pT$ from that of an
edge tile, and we can get $\cP_y\pR$ in a similar way from a ribbon
that is finite in $x$
instead.  However, unless we insist that the
bulk gauge is the same, we cannot use \eq{Qcsumd} to compute the
corner charge by summing these ingredients.
For example, if one obtains
$\cP_x\pT$ from a $y$-finite ribbon Wannierized along $\yhat$ and
$\cP_y\pR$ from an $x$-finite ribbon Wannierized along $\xhat$
as described in \sref{nw-trans} below,
then in general the gauges are not consistent,
and the sum $\cP_x\pT+\cP_y\pR$ is not meaningful.
(An exception to this rule will be discussed in \sref{2-band}.)

While preparing this manuscript, we became aware of
a recent work that proposes a ``thermodynamic'' definition of
gauge-invariant electric quadrupole
moments~\cite{daido-prb20}. However, the underlying
formulation of this approach is very different from ours;
it aims to describe local polarizations induced by
slow spatial variations of a bulk Hamiltonian, and makes
no claim to predict surface or corner properties except in
the case of quantizing symmetries. The two approaches
are thus complementary, and investigations into the relations
between them may be a fruitful avenue for future investigation.

\section{Methods}
\seclab{methods}

In this work, we use simple tight-binding models for the purpose
of implementing our formalism and testing its predictions.
These will be introduced in detail in \sref{results}.  Each model
is specified by providing the location of each basis orbital
$\ket{\varphi_{\0 i}}$ in
the rectangular $a\times b$ home unit cell, implying
periodic images
$\ket{\varphi_{\R i}}=T_\R\ket{\varphi_{\0 i}}$ in other cells.
The on-site energy of each basis orbital, and the hoppings
connecting near-neighbor orbitals, are also specified.
The position operator is assumed to be diagonal in the
tight-binding basis, $\me{\varphi_{\R i}}{\r}{\varphi_{\R'j}}=
(\R+\bm{\tau}_i)\delta_{\R,\R'}\delta_{ij}$,
with $\bm{\tau}_i$ denoting the location of the $i$th basis
function in the home cell. We treat the charge density
of each basis orbital as a Dirac delta function,
$|\ip{\r}{\varphi_{\R i}}|^2=\delta^2(\r-\R-\bm{\tau}_i)$,
so that the basis functions themselves have zero spread.
Positive ionic charges are assigned to all of
the tight-binding sites to neutralize the unit cell.
The electronic Hamiltonian
for bulk, ribbon, and flake geometries
is constructed and solved using the
\code{PythTB} code package~\cite{pythtb}.

\subsection{Corner charge and macroscopic averaging}
\seclab{macro-ave}

To calculate the corner charge directly, we construct a rectangular
flake consisting of $N_x\times N_y$ unit cells, and obtain the
total charge $q_{\R i}$ (ionic plus electronic) on every site.
  Since we associate the electronic charge to delta
functions on the sites, the total charge density takes the form
\beq
\rho(\r)=\sum_{\R i} q_{\R i} \delta^2(\r-\R-\bm{\tau}_i) \,.
\eqlab{rho-flake}
\eeq

The macroscopic corner charge is determined by
first applying a smoothening procedure, since simple sums of
individual charges are not convergent.  For this purpose we adopt
the sliding window average approach~\cite{resta-prl10,vanderbilt-book18},
in which a broadened charge density $\rhobar(\r)$ is obtained
by convoluting $\rho(\r)$ with a ``window function''
\beq
w(x,y)=
\begin{cases}
1/ab & \mbox{if }|x|<a/2,\;|y|<b/2 \\
0 & \mbox{otherwise}
\;,
\end{cases}
\eqlab{wxy}
\eeq
i.e.,
\beq
\rhobar(\r_0)=\int \rho(\r_0-\r')\,w(\r')\,d^2r' \,.
\eqlab{rhobaril}
\eeq
The advantage of this procedure is that $\rhobar(\r)$ is guaranteed to
vanish in the bulk-like regions of the sample as a result of the
charge neutrality of the bulk unit cell.
We also assume that the bulk has been terminated in such a
way as to yield neutral edges, as described in
\sref{quantized}, so that $\rhobar(\r)$ vanishes there as well.
The corner charge is then obtained by integrating the smoothened
charge density over the corner of interest.

\begin{figure}
\centering\includegraphics[width=2.6in]{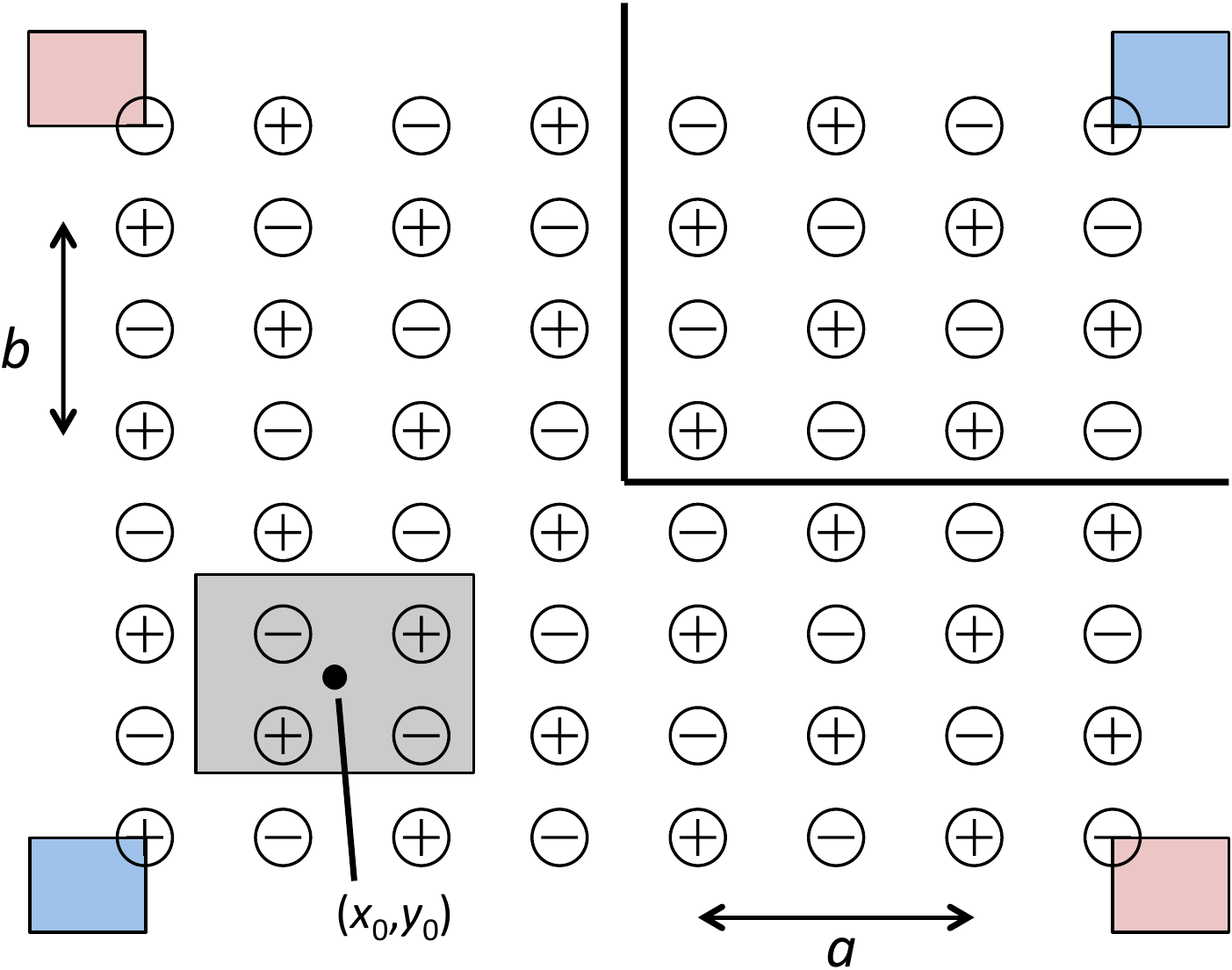}
\caption{Sketch of sliding-window approach for obtaining
macroscopic corner charges. The macroscopically averaged charge
density $\rhobar(\r_0)$ is defined as the average of $\rho(\r)$
over a rectangular cell centered at $\r_0=(x_0,y_0)$.
For this model of $\pm e$ point charges, $\rhobar$ vanishes except
in the blue and pink rectangles, where it takes values
$+e/ab$ and $-e/ab$ respectively.  Integration of the charge
in one of these colored rectangles yields the corresponding macroscopic
corner charge.}
\figlab{macro}
\end{figure}

The application of the above macroscopic averaging procedure to
a simple checkerboard arrangement of $\pm e$ point charges is
illustrated in \fref{macro}. The range of the window function centered
on position $\r_0=(x_0,y_0)$ is shown by the gray rectangle. As one
slides this window around on the sample, the charge contained in it
vanishes except when $\r_0$ falls in the rectangular $a/2\times b/2$
regions, where $\pm\rhobar=e/ab$ in the blue and red rectangles
respectively. Thus, the macroscopic edge charges are zero for this
model crystal, and the macroscopic corner charges are $\pm e/4$, with
the positive charges at top right and bottom left.

Other coarse-graining approaches will lead to the same result.
For example, Gaussian broadening can also be used, but then
a careful treatment of the two limits $\max(a,b) \ll \sigma \ll
\min(L_xa,L_yb)$ has to be enforced, where $\sigma$ is the
Gaussian width.  The sliding window approach
avoids such complications.  Note, however, that a simple summation
of the charges in a quadrant of the flake, as shown by the heavy
black lines, does \textit{not} yield the correct corner charge.
For the quadrant shown, it yields zero; if the quadrant contained
$5\times5$ instead of $4\times 4$ cells, it would yield $+e$.
Neither value is correct.

The technical implementation of the sliding window averaging
procedure is as follows. We can write
\beq
\Qc = \int d^2r\, \Lambda(\r) \, \rhobar(\r)
\eqlab{qcbar}
\eeq
where $\Lambda(\r)=1$ in an upper-right-hand region $x>x_0$
and $y>y_0$ and zero otherwise for appropriately chosen
$x_0$ and $y_0$.  In the language of function spaces this is
the inner product $\Lambda\circ\rhobar$, while
$\rhobar$ is the convolution $\rhobar=\rho*w$;
noting that $w(\r)=w(-\r)$,  this is equivalent
to $\Qc=W\circ\rho$ with $W=\Lambda*w$.  Thus,
in practice we compute the macroscopic corner charge as
\beq
\Qc = \int d^2r\, W(\r) \, \rho(\r) \,,
\eqlab{qcW}
\eeq
with $W(x,y)=f_a(x-x_0) f_b(y-y_0)$ given by the product of two
``ramp functions'' defined as $f_d(u)$~=~0 for $u<-d/2$,
$1$ for $u>d/2$, and $1/2+u/d$ in the interval $[-d/2,d/2]$.
Note that \eq{qcW} is not the same as the bare $\Qc$
obtained by integrating $\rho(\r)$ over a quadrant, i.e,
\beq
\Qc^{\rm bare} = \int d^2r\, \Lambda(\r) \, \rho(\r)
\eqlab{qcnom}
\eeq
for $x_0$ and $y_0$ at the sample center.  This definition of $Q_c$
was used in Refs.~[\onlinecite{benalcazar-prb17,vanmiert-prb18,
wheeler-prb19,kang-prb19}],
and the difference with respect to the macroscopic $Q_c$ of
\eqr{qcbar}{qcW} will be discussed in \sref{bbh}.

\subsection{Wannier construction for ribbon models}
\seclab{ribbons}

Our goal is to use our formalism to predict corner charges from
edge polarizations and interior quadrupoles computed for $x$-
and $y$-finite ribbon models.  For
example, we cut from the infinite 2D bulk a ribbon that is finite and
$N_y$ cells thick in the $y$ direction, but still infinite and
periodic in the $x$ direction.  In this case
the wavevector $k_x$ is a
good quantum number, and we obtain the Bloch states according to the
eigenvalue equation
\beq
H \ket{\psi_{k_xn}} = E_{k_xn} \ket{\psi_{k_xn}} \,.
\eqlab{Bloch}
\eeq
We are interested only in the occupied wavefunctions, so for
consistency with the bulk which has $J$ occupied bands, we let $n$ run
over $N_yJ$ occupied ribbon bands at each $k_x$.
We then need to construct a specific gauge for the
WFs spanning the occupied states, and in the following we present
three different strategies for doing so.

We first present, in \sref{proj}, a method based on projecting
onto trial functions.  As the same trial functions are used for
both $x$-finite and $y$-finite ribbons, this yields a consistent
gauge, allowing for a viable calculation of the corner charge.

Next, we discuss Wannier constructions based on assigning states to
layers via a preliminary maximal localization in one direction,
followed by maximal localization within each layer in the orthogonal
direction.  If the first step is taken in the transverse (finite)
direction, it corresponds to the ``hybrid Wannier'' construction; the
occupied subspace is represented in terms of states that are
exponentially localized in the transverse direction, while remaining
extended and labeled by wavevector in the longitudinal direction.
However, we then follow by a second localization step to arrive at
fully localized WFs.  This ``transverse-first'' nested Wannier
construction is described in \sref{nw-trans}.  We also consider the
reverse order of operations, in which the preliminary localization is
carried out in the extended direction; this ``longitudinal-first''
nested Wannier construction is described in \sref{nw-long}.

In Ref.~[\onlinecite{zhou-prb15}], the transverse-first hybrid Wannier
construction was applied to both the $x$-finite and $y$-finite
ribbons.  We emphasize that in general this does not produce the same
gauge for the interior WFs of the two ribbons, and hence it cannot
safely be used to predict the corner charge. This will later be
demonstrated explicitly in \sref{4-band}. (Centrosymmetric models with
a single occupied band and time-reversal symmetry provide an
exception, as will be discussed in \sref{2-band}.)  Instead, if either
the $x$-first or $y$-first nested Wannier scheme is consistently
adopted for \textit{both} ribbons (transverse for one ribbon and
longitudinal for the other), then we arrive at a second viable
approach for computing the corner charge, as discussed later in \sref{gcnw}.

In the following, we focus for concreteness on $y$-finite ribbons
and discuss each of the WF construction schemes in this context.

\subsubsection{Projection-based Wannier construction}
\seclab{proj}

One approach to the construction of a gauge, and one that
automatically produces the same gauge for both ribbons,
is to use the trial function projection
method~\cite{marzari-prb97,marzari-rmp12}. In this approach, one
invents $J$ trial functions $\ket{g_n}$ in the home unit cell
that are intended as a rough approximation to the desired bulk WFs, with
$g_{\ell_x\ell_yn}(\r)=g_n(\r-\ell_x\a-\ell_y\b)$ being their translational
images.  Then considering a $y$-finite ribbon, for example, we
construct a set of ribbon trial functions by taking the
$\ket{g_{\ell_x\ell_yn}}$ with
$\ell_x$ running over all integers while $\ell_y$ runs over
the $N_y$ layers in the ribbon, with possible additions or
deletions in the skin region to match the expected occupation
of edge and corner states (see, e.g., \sref{bbh}).
The goal then is to construct a set of WFs
$\ket{w_{\ell_x\ell_yn}}$ that look ``as similar as possible''
to these $\ket{g_{\ell_x\ell_yn}}$, while still being built only
from occupied Bloch states.

This is most easily done by going to reciprocal space.
Temporarily introducing the composite index $\alpha=(\ell_yn)$,
we define trial Bloch functions
\beq
\ket{\gg_{k_x\alpha}} = N_x^{-1/2}
  \sum_{\ell_x} e^{ik_x\ell_xa} \ket{g_{\ell_x\alpha}}
\eeq
and construct the overlap matrix
\beq
B_{k_x,\alpha\beta}=\ip{\psi_{k_x\alpha}}{\gg_{k_x\beta}}
\,.
\eeq
If our choice of trial functions had been ideal in the sense
that the $\ket{\gg_{k_x\alpha}}$ had spanned the occupied subspace
at $k_x$, $B_{k_x}$ would be a unitary matrix.  More
generally, we find the unitary part $\cB$ of the $B$ matrix
by subjecting it to the singular value decomposition
$B=V\Sigma W^\dagger$
($V$ and $W$ are unitary and $\Sigma$
is positive real diagonal), and choosing $\cB=VW^\dagger$.
We also monitor the singular values (diagonal elements of $\Sigma$);
if any of them becomes much less than unity, this signals the need
to choose a different set of trial functions.

We then construct mixtures of Bloch functions such that the resulting
ones are maximally aligned to the $\ket{\gg_{k_x\alpha}}$
according to
\beq
\ket{h_{k_x\alpha}}=\sum_\beta \cB_{k_x,\beta\alpha} \ket{\psi_{k_x\beta}}
\,.
\eeq
Restoring $\alpha=(\ell_yn)$, these $\ket{h_{k_x\ell_yn}}$ can be
interpreted as hybrid Wannier functions, as they are
exponentially localized in the finite direction while remaining
extended and labeled by wavevector $k_x$ in the extended direction.
From these, we can construct fully localized WFs by carrying
out the Fourier transform
\beq
\ket{w_{\ell_x\ell_yn}}=\frac{a}{2\pi}\int dk_x e^{-ik_x\ell_xa}
  \ket{h_{k_x\ell_yn}} \,.
\eqlab{ft-wan}
\eeq
In the deep interior of the ribbon, all
of these WFs will be periodic images of those in neighboring
cells.

We now pick the WFs associated with one
central cell with labels $(\ell_x\ell_y)$ and sum the
$\me{w_{\ell_x\ell_y n}}{xy}{w_{\ell_x\ell_y n}}$ over $n$
to obtain the interior quadrupole $q\bxy\pI$ via \eq{qxyI}, where
$\ket{\0 n}$ in the notation of \eq{qxyI} is the same as
$\ket{w_{\ell_x\ell_yn}}$ here.
Similarly,
we define the skin region at the top edge of the sample to
consist of some number $M_y$ of the top-most layers.
Since the dipole moments of
these cells vanish exponentially with depth, a fairly small value
of $M_y$ is typically sufficient.  Then, the $x$ dipole moments
$\me{w_{\ell_x\ell_y n}}{x}{w_{\ell_x\ell_y n}}$ are summed to
provide the needed contributions to the total dipole $d_x\pT$ of
\eq{dxf}.

We emphasize that our projection procedure insures that
if we start from the same set of trial
functions, the gauges in the interior region are the same by
construction for $y$-finite and $x$-finite ribbons.  Thus, we should
expect to find the same $\cQxy\pI$ for both
ribbons; we confirm this below.  Moreover, with the results of both
ribbon calculations in hand, we are assured that the set of quantities
$\cP\pT_x$, $\cP\pR_y$, and $\cQxy\pI$ have been
computed in a common gauge, and can confidently be combined as in
\eq{Qcsumd} to predict the corner charge.

\subsubsection{Transverse-first nested Wannier construction}
\seclab{nw-trans}

Let us now discuss an alternative Wannier construction procedure
that does not require choosing a set of trial functions. Again
taking a $y$-finite ribbon and noting that matrix elements of the
position operator $\yop$ are well defined, it is straightforward
to obtain the matrix
\beq
Y_{k_x,mn}=\me{\psi_{k_xm}}{y}{\psi_{k_xn}} \,,
\eqlab{Y-mat}
\eeq
where $m$ and $n$ run over the $N_yJ$ occupied bands of the
ribbon at a given $k_x$,
and to diagonalize it,
\beq
\sum_n Y_{k_x,mn} \xi_{k_x\alpha,n}=\ybar_{k_x\alpha}\xi_{k_x\alpha,m} \,,
\eqlab{Y-eig}
\eeq
where $\alpha=\{1,...,N_yJ\}$
now labels the eigenvalues and eigenvectors of $Y_{k_x}$.
Then the maximally localized states along $y$, known as
hybrid Wannier functions, are constructed according to
\beq
\ket{h_{k_x\alpha}}=\sum_n \xi_{k_x\alpha,n} \ket{\psi_{k_xn}} \,.
\eqlab{Y-rot}
\eeq
As we shall see, the spatial locations of their Wannier centers
$\ybar_{k_x\alpha}$ cluster in groups of $J$ per unit cell along $y$,
corresponding roughly to the locations along $y$ of the true 2D
WFs assigned to a unit cell.  Thus, we relabel
$\ybar_{k_x\alpha} \rightarrow \ybar_{k_x\ell_y n}$ and
$\ket{h_{k_x\alpha}} \rightarrow \ket{h_{k_x\ell_y n}}$, where
$\ell_y$ is a layer index specifying the unit cell along $y$ and
$n=\{1,...,J\}$ labels the Wannier bands
within a layer.

Then, for each layer that has been identified in this way,
we treat the entire layer as a multiband group, and carry out
a maximal localization procedure in the extended direction.
To do so, we transform to a
twisted parallel transport gauge, i.e., one that makes the
the Berry connections $\me{\tilde{h}_{k_x\ell_yn}}
{i\partial_{k_x}} {\tilde{h}_{k_x\ell_yn'}}$ diagonal and $k_x$-independent,
where $\ket{\tilde{h}_{k_x\ell_yn}} = e^{-ik_xx}\ket{h_{k_x\ell_yn}}$.
The fully localized WFs are constructed from the Fourier
transform in \eq{ft-wan}, thus arriving at WFs that are
exponentially localized in both directions. The computation of
$q\bxy\pI$ from deep interior WFs, and $d_x\pT$ from skin-region
WFs, then proceeds as described in the previous subsection.

We note in passing that another option for computing $d\pT_x$ is to
bypass the second maximal localization step and simply compute
it from Berry phases, as was done in Ref.~\cite{zhou-prb15}.
That is, having constructed the $\ket{\tilde{h}_{k_x\ell_yn}}$, we
compute the Berry phases
\beq
\gamma^{(x)}_{\ell_y n}=\int dk_x\, \me{\tilde{h}_{k_x\ell_yn}}
{i\partial_{k_x}} {\tilde{h}_{k_x\ell_yn}}
\eqlab{bpcalc}
\eeq
on a discretized $k_x$ mesh
using standard methods.
In this context the last term in \eq{dxf} becomes
$(-e/2\pi) \sum_{\ell_yn}\pT \gamma^{(x)}_{\ell_y n}$,
where the sums are restricted to the cells associated with the
top-edge tiles.
However, we find in practice that $P\pT_x$ computed in this way
converges more slowly with respect to $k$-mesh density
than does the method based on the direct summation of WF dipoles,
which we have therefore adopted below.

\subsubsection{Longitudinal-first nested Wannier construction}
\seclab{nw-long}

The nested procedure outlined in the previous subsection consists of
a sequence of two maximal localization steps,
the first along the ribbon's finite direction $y$ and the second
along the extended direction $x$. If we reverse the order of those two
operations, we again arrive at fully localized WFs, albeit in a
different gauge. Since the first localization step is now along the
extensive direction of the ribbon, we refer to this as the
longitudinal-first nested Wannier construction.
We note that a similar construction was used in
Refs.~\cite{benalcazar-sa17,benalcazar-prb17},
although the subsequent steps making use of the construction were
different there.

We again start from the Bloch eigenstates $\ket{\psi_{k_xn}}$ of
\eq{Bloch}. We first transform all of them to a twisted parallel transport
gauge in the extensive direction $x$,
and then carry out the Fourier transform
\beq
\ket{h'_{\ell_x n}}=\frac{a}{2\pi}\int dk_x e^{-ik_x\ell_xa}
  \ket{\psi_{k_xn}}\,.
\eqlab{ft-hw}
\eeq
These new states
are maximally localized along $x$, but typically they are extended
across the width of the ribbon in the $y$ direction.
In a sense, they can still be regarded as a species of hybrid WFs.
Those with the same index $n$ but
different cell indices $\ell_x$ are translational copies of one
another along $x$.  Finally we localize along $y$ the $N_yJ$ hybrid Wannier
functions in each horizontal cell $\ell_x$ by performing the steps in
\eqr{Y-mat}{Y-rot} with $\ket{\psi_{k_xn}}$ therein replaced by
$\ket{h'_{\ell_x n}}$. This yields a set of fully localized WFs
$\ket{w_{\ell_x\ell_yn}}$, from which the
interior quadrupole $q\pI\bxy$ and
edge dipoles $d\pT_x$
can be evaluated as described below \eq{ft-wan}.

\subsubsection{Quantum distance between Wannier gauges}
\seclab{distance}

Once specific gauges have been chosen for
differently oriented ribbons or different Wannier constructions,
it is useful to check whether those gauges are consistent.
By ``consistent gauges'' we mean that the sets
$\{\vert w_{{\rm int},n}\rangle\}$ and
$\{\vert \widetilde{w}_{{\rm int},n}\rangle\}$
of $J$ WFs in one interior cell span the same Hilbert space in both
cases.  If so, the two sets of WFs are related by a $J\times J$
unitary transformation
\beq
\ket{\widetilde{w}_{{\rm int},n}}=\sum_{m=1}^J\, U_{mn}\ket{w_{{\rm int},m}}
\eqlab{U-intra}
\eeq
that only mixes WFs within the same interior cell. On the other
hand, \eqr{Qvar}{Pyvar} show that $\cQxy\pI$,
$\cP_x\pT$, and $\cP_y\pR$ only change under gauge transformations
that mix WFs belonging to different cells ($\R'\not= {\bf 0}$). This
means that we are allowed to evaluated the corner charge as the sum of
those three quantities provided that they are evaluated using
gauges for the two ribbons
that are consistent in the above sense.

The degree of ``gauge inconsistency'' can be quantified by
measuring the ``quantum distance'' between the two sets of interior
WFs. Here the square of the quantum distance $D$ is defined
as~\cite{liu-prb14}
%
\begin{align}
D^2&=J-\Tr\left[\cP_{\rm int} \widetilde\cP_{\rm int}\right]\nn
&=J-\sum_{m,n=1}^J\,
\left|\ip{w_{{\rm int},m}}{\widetilde w_{{\rm int},n}}\right|^2\,,
\eqlab{dist}
\end{align}
where $\cP_{\rm int}$ and $\widetilde \cP_{\rm int}$ are the projection
operators onto each set. A vanishing $D$ indicates that the two sets
are related by a unitary transformation.  Allowing for numerical
error, we take the gauges to be consistent whenever $D<10^{-5}$.

\section{Results}
\seclab{results}

We study three tight-binding models of increasing complexity.
All models are centrosymmetric and spinless, and we consider them at
half filling.  The first is a two-band model (one occupied band), and
the other two are four-band models (two occupied bands). In the first
two models the symmetry is sufficiently low that the corner charge is
not quantized, while the third model has a high-symmetry phase where
the corner charge is quantized to either zero or $e/2$, depending on
the choice of parameters.
For ribbons and finite flakes, edges are always constructed
by simply truncating the bulk, i.e., the hoppings to vacant sites
are removed while other hoppings and site energies are unchanged.

In this section, we restrict ourselves to a comparison of the
transverse-first nested Wannier construction as applied to both
ribbons, as in Ref.~[\onlinecite{zhou-prb15}], and the projection
construction.  In \sref{gcnw} we will return to the four-band model
of \sref{4-band} and consider
the gauge-consistent nested Wannier construction,
i.e., $y$-first (or $x$-first) for both ribbons,
and show that this also yields a consistent gauge
and a correct prediction of the corner charge.

\subsection{Two-band model}
\seclab{2-band}

\begin{figure}
\centering\includegraphics[width=3.3in]{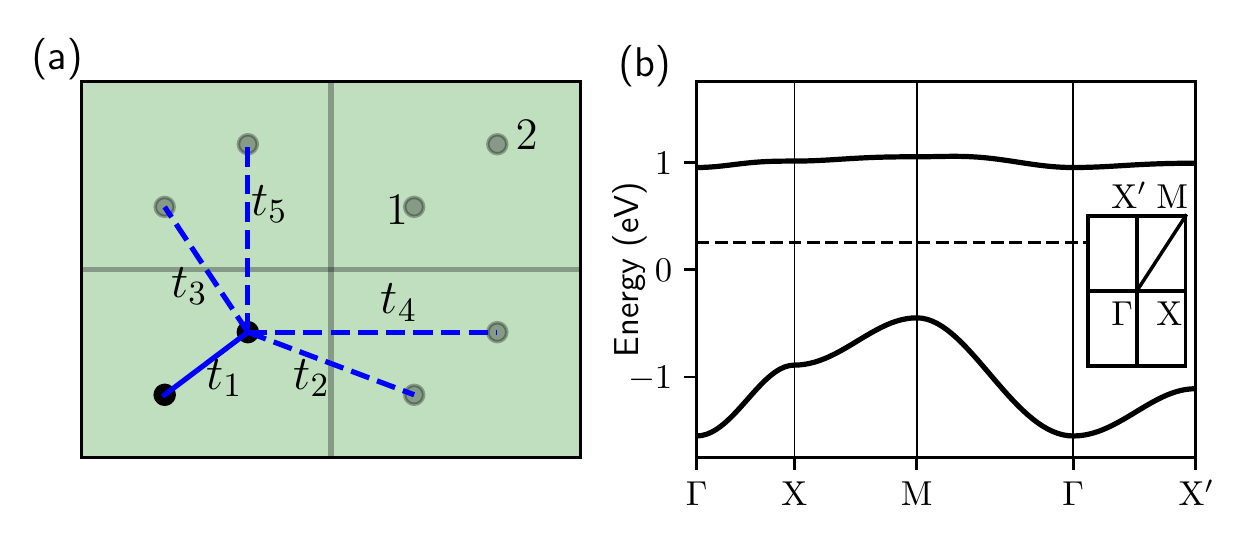}
\caption{(a) Visualization of the two-band model. Atoms in the home
  cell are shown as black dots, the intracell hopping $t_1$ as a solid
  line, and intercell hoppings $t_2$ to $t_5$ as dashed lines. (b)
  Bulk band structure for the parameters given in the main text. The
  Fermi level (dashed line) has been placed at midgap.
  The inset shows the 2D Brillouin zone and the high-symmetry points
  $\Gamma$ $(0,0)$, ${\rm X}$ $(\frac{1}{2},0)$, ${\rm X}^\prime$
  $(0,\frac{1}{2})$, and ${\rm M}$ $(\frac{1}{2},\frac{1}{2})$.}
  \figlab{2-band-model}
\end{figure}

The first model we consider was introduced in
Ref.~[\onlinecite{zhou-prb15}], and is illustrated in
\fref{2-band-model}(a). The rectangular unit cell (gray square)
has an aspect ratio
of $b/a=0.8$, and contains two atoms along its diagonal, with reduced
coordinates $(-\frac{1}{6},-\frac{1}{6})$ and
$(+\frac{1}{6},+\frac{1}{6})$ relative to the center of inversion in
the middle of the cell.  Since we treat
the model as spinless and at half filling, we assign a positive
charge of $+e/2$ to each atom to neutralize the unit cell.

Our choice of bulk tile corresponds to the contents of the unit cell
in \fref{2-band-model}(a), with the reference position $\bm{\tref}_n$ chosen
at the origin, which is also the location of the WF center.  As a
result, the ionic part of the interior quadrupole $q\bxy\pI$ of
\eqr{qxyI}{qxyif} is immediately given as $q\bxy\pion=(e/36)ab$.  The
electronic contribution $q\bxy\pel$ in \eq{qxyef} is determined by the
shape asymmetry of the WF charge distribution around its center, and
remains to be calculated, as do the dipoles of the edge tiles.  From
these, $\cQxy\pI$, $\cP_x\pT$, and $\cP_y\pR$ are trivially
obtained from \eqr{QxyI}{Qcsumd}.

To evaluate these quantities
we construct two ribbons spanning ten unit cells
along the $x$ and $y$ directions respectively.
We begin by applying the transverse-first nested Wannier
construction of \sref{nw-trans} to both ribbons.  That is, the maximal
localization procedure is first carried out along the finite direction
of the ribbon to generate hybrid WFs, and then along the
extended direction. The result is illustrated in \fref{2-band-hwf}
for the $y$-finite ribbon. Panel~(a) shows the Wannier centers
$\bar{y}_{k_x\ell_y}=\me{h_{k_x\ell_y}}{y}{h_{k_x\ell_y}}$ obtained in
the first step.  In the second step, an optimally-smooth gauge
along $x$ is enforced within each hybrid Wannier band, resulting in
fully localized WFs.
Panel~(b) shows the layer-resolved dipole moment
density along $x$; as expected, it vanishes in the interior region and
assumes equal and opposite values at the two edges.

\begin{figure}
\centering\includegraphics[width=3.3in]{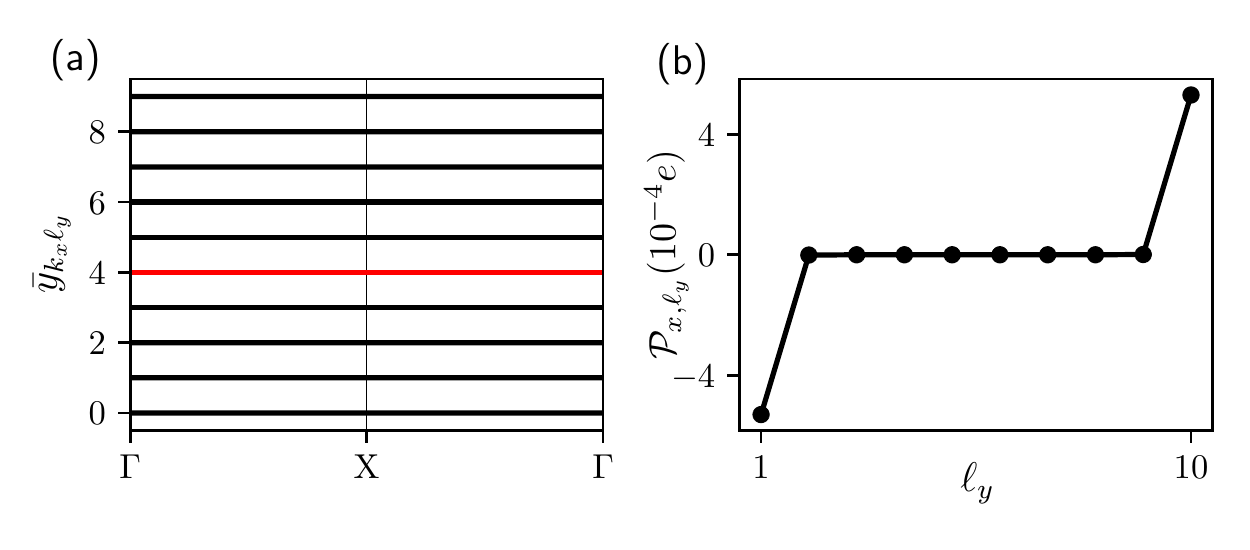}
\caption{(a) Wannier bands (i.e, hybrid Wannier centers)
  $\bar{y}_{k_x\ell_y}$ (in units of $b$) for a ribbon of the
  two-band model with a width of ten unit cells along $y$. The red
  Wannier band deep inside the ribbon is selected to construct the
  fully localized interior Wannier function that is
  used to evaluate $\cQxy\pI$. (b) Layer-resolved dipole density
  $\cP_{x,\ell_y}=d_{x,\ell_y}/a$, computed from the dipole moments of
  the fully localized Wannier functions in each layer.}
\figlab{2-band-hwf}
\end{figure}

\begin{table}[b]
\caption{\label{tab:2-band-result} The values of $\cP_x\pT$,
  $\cP_y\pR$, and $\cQxy\pI$, calculated
  for ribbons of the two-band model using the
  transverse-first hybrid Wannier and
  projection methods. In the bottom half of the table, the corner
  charge $\Qc$ predicted from \eq{Qcsumd} is compared with the value
  obtained from a direct calculation on a finite flake using \eq{qcW},
  and with the ``bare'' corner charge obtained from \eq{qcnom}.}
\begin{ruledtabular}
\begin{tabular}{lcc}
 & Hybrid Wannier  & Projection \\
 & ($10^{-3}~e$) & ($10^{-3}~e$)\\
\colrule
$\cP_x\pT$  & $-0.531575$ & $-0.531574$   \\
$\cP_y\pR$ & $-1.164427$ & $-1.164427$    \\
$\cQxy\pI$ & $-0.030068$ & $-0.030068$  \\
\colrule$\Qc$ (predicted) & $-1.726070$ & $-1.726069$ \\
$\Qc$ (direct) & $-1.726068$ & $-1.726068$ \\
$\Qc^{\rm bare}$ & $-0.071873$ & $-0.071873$
\end{tabular}
\end{ruledtabular}
\end{table}

The values of $\cP_x\pT$, $\cP_y\pR$, 
and $\cQxy\pI$ calculated from
those WFs are indicated in the left column of
Table~\ref{tab:2-band-result}. We find that $\cQxy\pI$ has the same
value in the two ribbons, suggesting that their gauges are
consistent.
Decomposing $\cQxy\pI$ into ionic and electronic parts, we find
$\cQxy\pion = q\bxy\pion/ab = e/36\approx 0.027778e$ and
$\cQxy\pel = q\bxy\pel /ab = -0.027808e$.
We also find that the corner charge predicted from
\eq{Qcsumd} is in excellent agreement with that obtained from a direct
calculation on a $10\times 10$ flake using \eq{qcW}, again suggesting
that the gauges are consistent (as well as validating our
formalism). The last row of Table~\ref{tab:2-band-result} lists the
value of the bare corner charge, obtained by simply adding up the
charges inside the $5\times 5$ tiles forming the top-right quadrant
of the flake, according to \eq{qcnom}; as expected,
the bare corner charge differs significantly from the macroscopic
corner charge listed in the two rows above it.

\begin{table}
\caption{\label{tab:2-band-bulk-wf} The bulk-like Wannier function
  $\ket{w_{\rm int}}$ in the home unit cell of the two-band model.
  $\ket{\phi_{\R j}}$ is the basis orbital at site $\R+\taub_j$, given
  in reduced coordinates. The 12 largest coefficients are listed;
  only half of them are shown, as the other half can be obtained
  by an inversion operation.}
\begin{ruledtabular}
\begin{tabular}{cccc}
$\R+\taub_j$ & $\ip{\phi_{\R j}}{w_{\rm int}}$ & $\R+\taub_j$ & $\ip{\phi_{\R j}}{w_{\rm int}}$\\
\colrule
($-\frac{1}{6},-\frac{1}{6}$)  & $\phm0.70565$ & $(-\frac{5}{6},\frac{1}{6})$ & $0.02662$   \\
$(-\frac{7}{6},-\frac{1}{6})$ & $-0.02634$ & $(-\frac{1}{6},\frac{5}{6})$ & $0.01777$   \\
$(-\frac{1}{6},-\frac{7}{6})$ & $-0.01752$ & $(\phm\frac{7}{6},\frac{7}{6})$ & $0.00328$ \\
\end{tabular}
\end{ruledtabular}
\end{table}

To confirm that the gauges are consistent between the two ribbons, we
calculate the quantum distance~$D$ according to \eq{dist}, and find
that it is zero to numerical accuracy.
Since there is a single WF per cell, gauge consistency means that the
WFs deep inside the two ribbons are the same up to an overall phase
factor. The site amplitudes of one such interior WF are listed in
Table~\ref{tab:2-band-bulk-wf}.

Recall that the transverse-first nested Wannier construction is not
guaranteed to yield consistent gauges for two differently oriented
ribbons of a generic model. The reason why it does so for this
particular model is the following. In addition to spatial inversion,
the model has time-reversal symmetry, and in the presence of both
symmetries the $k$-space Berry curvature of each band vanishes
identically.  Since the curvature is the curl of the connection, it
follows that both the $x$ and $y$ components of the Berry connection
can be chosen to be constant.  Moreover, these constant values are a
measure of the electric polarization, which vanishes here.  Thus, in
this case of a single occupied band with inversion and time-reversal
symmetry, there is a unique ``natural'' gauge with vanishing Berry
connection.  This same gauge is arrived at regardless of whether
maximal localization is applied first in $x$ and then in $y$, first in
$y$ then in $x$, jointly as in conventional 2D maximal localization,
or using the projection technique discussed next.%
\footnote{The order of the two Wannierization steps becomes irrelevant
  when the projected position operators $PxP$ and $PyP$ commute,
  which was shown in Appendix C of
  Ref.~[\onlinecite{marzari-prb97}] to occur if and only if
  the Berry curvature vanishes identically.}

We now repeat the calculations using the projection method of
\sref{proj} to fix the gauge. We choose as the trial function the
eigenstate of an isolated tile, without any inter-cell hoppings. The
trial function in the home unit cell is then
$\frac{1}{\sqrt{2}}\ket{\phi_1}+\frac{1}{\sqrt{2}}\ket{\phi_2}$, where
$\ket{\phi_1}$ and $\ket{\phi_2}$ are the basis orbitals located at
$(-\frac{1}{6},-\frac{1}{6})$ and $(\frac{1}{6},\frac{1}{6})$,
respectively.  After confirming that the resulting gauges for the two
ribbons are consistent ($D=0$ to numerical accuracy),
we have recalculated $\cP_x\pT$,
$\cP_y\pR$, and $\cQxy\pI$, obtaining the values in the right column
of Table~\ref{tab:2-band-result}. They are identical to the ones in
the left column, confirming that the transverse-first hybrid Wannier
and projection methods yield
consistent gauges for this model. To further verify this, we measure
the quantum distance between the interior WFs obtained with the two
methods, again obtaining $D=0$. 

We conclude by commenting on the results obtained in
Ref.~[\onlinecite{zhou-prb15}] for the same model. In that work,
$\cP_x\pT$ and $\cP_y\pR$ were calculated for $y$- and $x$-finite
ribbons using the transverse-first nested Wannier construction,
and $\cP_x\pT+\cP_y\pR$ was found to be in good agreement with a
direct calculation of $\Qc$ for a flake. Our analysis reveals an
oversight in that work, also pointed out in
Ref.~[\onlinecite{trifunovic-prresearch20}], namely the omission of
the $\cQxy\pI$ term in \eq{Qcsumd}.  For the choice of parameters in
Ref.~[\onlinecite{zhou-prb15}], $|\cQxy\pI|$ is much smaller than both
$|\cP_x\pT|$ and $|\cP_y\pR|$, helping to explain why that omission
was not revealed by the numerical tests carried out there.
Reference~[\onlinecite{zhou-prb15}] also neglected to discuss the
gauge-consistency issue that arises in more general cases, although as
discussed above, it is not a problem for single-occupied-band models
with time-reversal symmetry.  It does become an issue for multiband
cases, as we shall see in our next example.

\begin{figure}
\centering\includegraphics[width=3.3in]{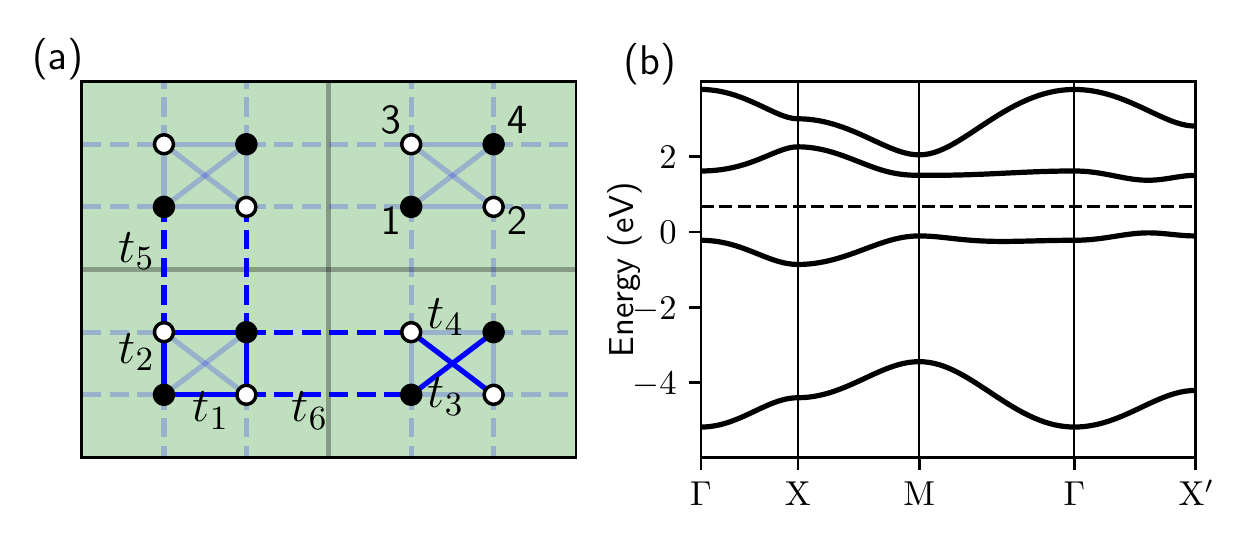}
\caption{(a) Visualization of the four-band model. Atoms are labeled
  from 1 to 4 as shown at upper right.  The intracell hoppings $t_1$,
  $t_2$, $t_3$ and $t_4$ are shown as solid lines, while the intercell
  hoppings $t_5$ and $t_6$ are shown as dashed lines. Sites denoted by
  open and filled circles have on-site energies $\pm\delta$
  respectively. (b) Band structure of the model. The Fermi energy
  (dashed line) has been placed at midgap.}
\figlab{4-band-model}
\end{figure}

\subsection{Four-band model}
\seclab{4-band}

Our second test case is the model depicted in
\fref{4-band-model}(a). The unit cell is rectangular with $b/a=0.8$
as before, but it now contains four atoms instead of two,
with reduced coordinates
$(-\frac{1}{6},-\frac{1}{6})$, $(\frac{1}{6},-\frac{1}{6})$,
$(\frac{1}{6},\frac{1}{6})$ and $(-\frac{1}{6},\frac{1}{6})$
relative to the center of inversion in the middle of the
cell. The hopping amplitudes are $t_1 = -2.0$, $t_2 = -1.5$,
$t_3 = -0.8$, and $t_4 = -0.6$~eV (intracell hoppings), and
$t_5=-0.5$ and $t_6=-0.4$~eV (intercell hoppings).
The sites depicted as open and filled circles have onsite energy
$\pm\delta$, where $\delta = 0.8$. The band structure is shown in
\fref{4-band-model}(b); at half filling the two lowest bands are
occupied, and we assign a charge of $+e/2$ to each atom to render the
cell neutral.  The bulk tile again corresponds to the unit cell, and
the reference positions of \eqr{rhoion}{rhoel} are again
$\bm{\tref}_1=\bm{\tref}_2=\0$; now $q\bxy\pion=0$ and only $q\bxy\pel$ will
contribute to $q\bxy\pI$.

\begin{figure}
\centering\includegraphics[width=\columnwidth]{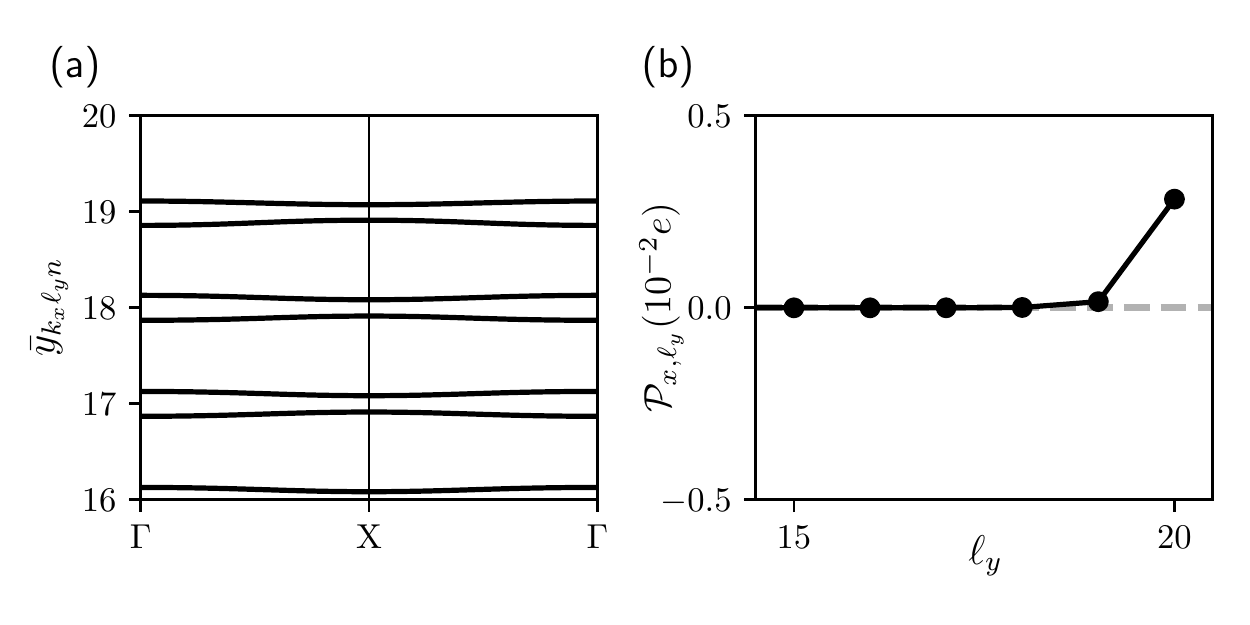}
\caption{(a) Wannier bands $\bar{y}_{k_x\ell_yn}$ (in units of $b$)
  for a $y$-finite ribbon of the four-band model with a width
  of 20 layers, and two bands per layer.  (b) Dipole moment density of
  the layers near the top edge of the ribbon.}
\figlab{4-band-hwf}
\end{figure}

As in our previous example, the model has both spatial inversion and
time-reversal symmetry.  However, since we now have two occupied
bands, the transverse-first nested Wannier construction is no
longer expected to produce consistent gauges for the two ribbons.
Its application to a 20-cell-thick $y$-finite ribbon
is illustrated in
\fref{4-band-hwf}. Panel~(a) shows the Wannier bands obtained in the
first step, with the two bands in each vertical cell being closer to
one another than to their neighbors in adjacent cells. In the second
step, the maximal localization procedure is applied along $x$,
treating the two hybrid Wannier functions within a cell as a
composite group, resulting in two fully localized WFs per 2D cell.
When applied to the $x$-finite ribbon, the transverse-first
procedure results in a similar pair of WFs, but now obtained
by localizing first along $x$ and then along $y$.

\begin{table}[b]
\caption{\label{tab:4-band-result} The values of $\cP_x\pT$,
  $\cP_y\pR$, $\cQxy\pI$, calculated for ribbons of the four-band
  model using the transverse-first hybrid Wannier and projection
  methods. In the center column, the identical value of
  $\cQxy\pI$ found for both ribbons is reported.
  In the last two rows, the corner charge $\Qc$ predicted from
  \eq{Qcsumd} is compared with the value obtained from a direct
  calculation on a finite flake.}
\begin{ruledtabular}
\begin{tabular}{lcc}
&  Hybrid Wannier & Projection \\
&  $(10^{-2}~e)$ & $(10^{-2}~e)$ \\
\colrule
$\cP_x\pT$ & $\phm0.300250$ & $\phm0.254669$   \\
$\cP_y\pR$ & $\phm0.476420$ & $\phm0.446029$   \\
$\cQxy\pI$ & $-3.756016$ & $-3.684265$   \\
\colrule
$\Qc$ (predicted) & $-2.979346$ & $-2.983567$  \\
$\Qc$ (direct) & $-2.983567$ & $-2.983567$
\end{tabular}
\end{ruledtabular}
\end{table}

The center column of Table~\ref{tab:4-band-result} lists the
calculated values of $\cP_x\pT$ (for the $y$-finite ribbon),
$\cP_y\pR$ (for the $x$-finite ribbon), and $\cQxy\pI$ (for both).
Even though $\cQxy\pI$ has identical values in both ribbons,
the predicted corner charge $\Qc$ differs by about 0.14\% from that
obtained via a direct calculation on a $20\times 20$ flake, indicating
some degree of gauge inconsistency.  The gauges of the two ribbons are
indeed slightly different, as can be seen by inspecting the second and
third columns of Table~\ref{tab:4-band-bulk-wf}, where we list the
site amplitudes of one of the two interior WFs per cell (the other is
related to it by spatial inversion) in each ribbon.  To check that
this difference cannot be accounted for by a $2\times 2$ intracell
gauge transformation described by \eq{U-intra}, we calculate the
quantum distance of \eq{dist} to be $D=0.0138$.  This nonzero value
confirms that the interior gauges produced by this naive
hybrid Wannier approach is
inconsistent between the two ribbons.

To arrive at a common gauge for the two ribbons we use the
projection method, choosing as trial functions
$\ket{g_1} =
\frac{1}{\sqrt{2}}\ket{\psi_1}+\frac{1}{\sqrt{2}}\ket{\psi_2}$ and
$\ket{g_2} =
\frac{1}{\sqrt{2}}\ket{\psi_1}-\frac{1}{\sqrt{2}}\ket{\psi_2}$,
where $\ket{\psi_1}$ and $\ket{\psi_2}$ are the two lowest-energy
eigenstates of an isolated tile, i.e., with intercell hoppings set
to zero.  These two eigenstates are of even
and odd parity respectively,
so that $\ket{g_1}$ and $\ket{g_2}$ are each off-centered with
respect to the origin, and map into one another under inversion.

\begin{table}
\caption{\label{tab:4-band-bulk-wf} One of the two bulk-like Wannier
  functions in the home unit cell of the four-band model, constructed
  in three different ways.  $\ket{w_{{\rm int},1}^{(y)}}$ and
  $\ket{w_{{\rm int},1}^{(x)}}$ are obtained by applying the
  transverse-first nested Wannier construction
  to $y$- and $x$-finite ribbons, respectively, while
  $\ket{w_{{\rm int},1}^{({\rm p})}}$ is obtained by applying the
  projection method to both ribbons starting from the trial
  function $\ket{g_{1}}$ described in the main text. $\ket{\phi_{\R j}}$ is the
  basis orbital at site $\R+\taub_j$, given in reduced coordinates.}
\begin{ruledtabular}
\begin{tabular}{ccccc}
$\R+\taub_j$ & $\ip{\phi_{\R j}}{w_{{\rm int},1}^{(y)}}$ &
$\ip{\phi_{\R j}}{w_{{\rm int},1}^{(x)}}$ & $\ip{\phi_{\R j}}{w_{{\rm int},1}^{({\rm p})}}$  & $\ip{\phi_{\R j}}{g_{1}}$\\
\colrule
$(-\frac{1}{6},-\frac{1}{6})$  & $-0.86557$ & $-0.86563$ & $-0.86481$ & $-0.87128$ \\
$(-\frac{1}{6},\phm\frac{1}{6})$ & $-0.42664$ & $-0.42656$ & $-0.42851$ & $-0.45897$ \\
$(\phm\frac{1}{6},-\frac{1}{6})$ & $-0.17659$ & $-0.17654$ & $-0.17732$ & $-0.15379$ \\
$(-\frac{1}{6},-\frac{5}{6})$ & $-0.11485$ & $-0.11527$ & $-0.10720$ & $0$ \\
$(-\frac{5}{6},-\frac{1}{6})$ & $-0.07271$ & $-0.07294$ & $-0.07163$ & $0$\\
$(\phm\frac{1}{6},\phm\frac{1}{6})$  & $\phm0.07108$ & $\phm0.07108$ & $\phm0.07185$ & $\phm0.08101$\\
$(\phm\frac{1}{6},\phm\frac{5}{6})$  & $\phm0.06598$ & $\phm0.06548$ & $\phm0.06286$ & $0$\\
$(\phm\frac{5}{6},\phm\frac{1}{6})$  & $\phm0.04861$ & $\phm0.04798$ & $\phm0.04626$ & $0$\\
\end{tabular}
\end{ruledtabular}
\end{table}

Applying the projection method to ribbon models cut from the bulk
as described in \sref{proj}, we find as expected that the pair of
WFs taken from the deep interior of the $x$-finite ribbon
match those extracted from the $y$-finite ribbon within numerical
precision.
We denote as $\ket{w_{{\rm int},1}^{({\rm p})}}$ and
$\ket{w_{{\rm int},2}^{({\rm p})}}$ the WFs projected from $\ket{g_1}$
and $\ket{g_2}$ respectively. Like the trial functions, these lie
off-center and map into one another under inversion.  In the last two
columns of Table~\ref{tab:4-band-bulk-wf} we list the site amplitudes
of $\ket{w_{{\rm int},1}^{({\rm p})}}$ and $\ket{g_1}$.  It is
evident that the projected WFs are similar, but not identical, to
the ones obtained by the transverse-first hybrid
Wannier approach; we find a quantum distances of
$D=0.03844$ and $0.03857$ respectively from the projected pair to the
pairs generated via the transverse-first nested Wannierization
of $x$-finite and $y$-finite ribbons respectively.

Having verified that the projection method leads to
two ribbons described by the same bulk gauge, we
proceeded to calculate $\cP_x\pT$ for the $y$-finite ribbon and
$\cP_y\pR$ for the $x$-finite ribbon; their values are listed in the
right column of Table~\ref{tab:4-band-result}, followed by the common
value of $\cQxy\pI$ in both ribbons. In contrast to the center column,
the sum of the three now matches perfectly the value of $\Qc$ in the
finite flake.

This example confirms our expectation that the corner charge can
reliably be predicted from ribbon calculations alone, provided that
consistent gauges are used for both ribbons, even in the case of
multiple occupied bands. It also illustrates the fact that this gauge
consistency is achieved only via the projection method,\footnote{A
  gauge-fixing method was recently proposed in
  Ref.~[\onlinecite{trifunovic-prresearch20}] based on parallel
  transport as intercell hoppings are varied.
  Although this was applied only to a single-occupied-band
  case, we expect that this method, while more complicated than
  ours, would also lead to bulk-like WFs in a multi-band case.}
while the transverse-first hybrid Wannier approach fails in this case.

\subsection{Benalcazar-Bernevig-Hughes (BBH) model}
\seclab{bbh}

\begin{figure}
\centering\includegraphics[width=\columnwidth]{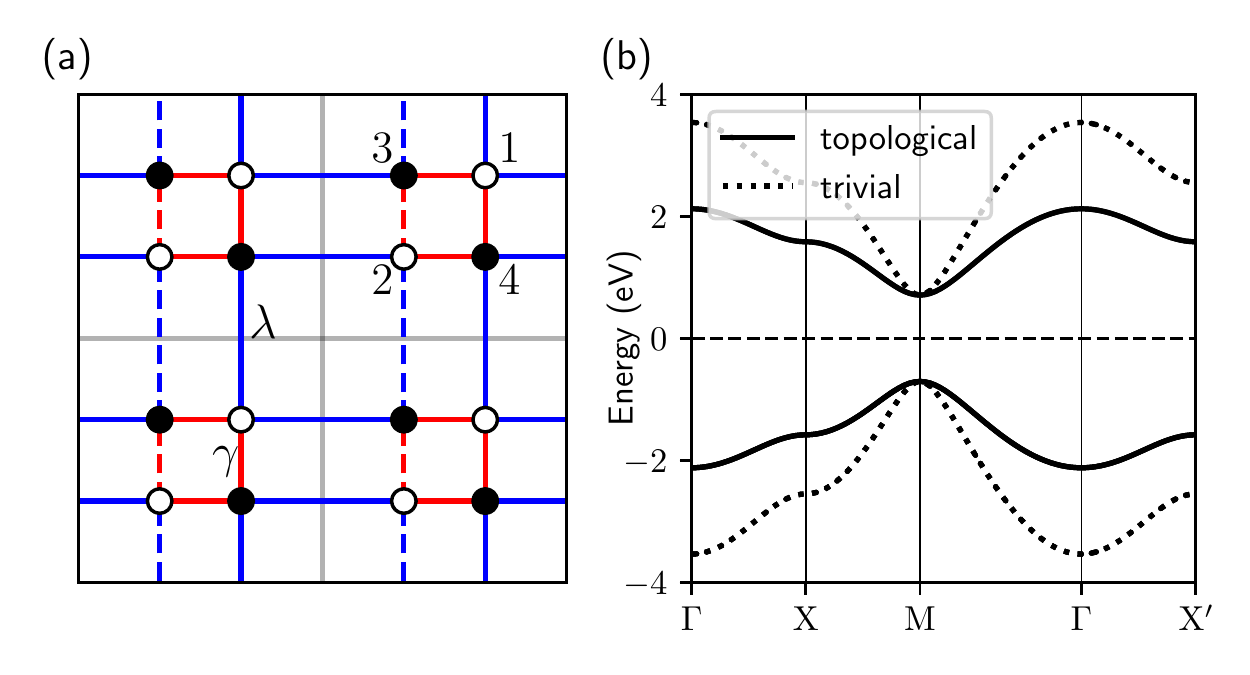}
\caption{(a) Visualization of the Benalcazar-Bernevig-Hughes
  model. Atoms are labeled from 1 to 4 inside the unit cell, as shown
  at upper right. Intracell hoppings of amplitude $\pm\gamma$ are
  shown as solid or dashed red (short) lines, and intercell hoppings
  of amplitude $\pm\lambda$ as solid or dashed blue (long) lines. In
  the calculations reported in the main text, sites denoted by
  open and filled circles have small on-site energies
  $\pm\delta$ respectively. (b) Band structure of the model for
  $\lambda=1.0$, $\delta=0$, and two different values of $\gamma$.
  The solid bands were calculated at $\gamma=1.5$ (trivial phase,
  $\Qc=0$), and the dashed ones at $\gamma=0.5$ (topological phase,
  $\Qc=e/2$). In both cases, the bands are doubly degenerate. The
  Fermi energy (dashed line) has been placed at midgap.}
\figlab{bbh-band-model}
\end{figure}

Our final test case is a model introduced by Benalcazar, Bernevig, and
Hughes as an example of a topological phase with quantized corner
charges~\cite{benalcazar-sa17,benalcazar-prb17}. The BBH model is
pictured in \fref{bbh-band-model}(a). It has four sites per cell as in
our previous example, but now placed on a square lattice.  We again
choose the atoms to have reduced coordinates
$(-\frac{1}{6},-\frac{1}{6})$, $(\frac{1}{6},-\frac{1}{6})$,
$(\frac{1}{6},\frac{1}{6})$ and $(-\frac{1}{6},\frac{1}{6})$ relative
to the origin at the center of a small square.\footnote{The
  location of the sites was not specified in
  Refs.~[\onlinecite{benalcazar-sa17,benalcazar-prb17}].
  Our choice of 1/6 is arbitrary.}
\Fref{bbh-band-model}(a) shows four unit cells (gray squares)
centered in the same way, but as we shall see later, our choice
of bulk tile may or may not coincide with this unit cell.
Each site also carries an ionic charge of $+e/2$, so that the
system is neutral at half filling.

When viewed along $x$ or $y$, the model consists of parallel chains
with dimerized bonds. The hopping amplitudes along $x$ alternate
between $\gamma$ (intracell) and $\lambda$ (intercell). The same bond
alternation occurs along $y$, except that the hopping amplitudes
change sign from one chain to the next, as though $\pi$ fluxes have
been threaded through the plaquettes.  Following BBH, we also include
an optional parameter $\delta$ which, if present, assigns an on-site
energy $\pm\delta$ to the sites depicted as
open and filled circles respectively
in \fref{bbh-band-model}(a).

The model always has inversion and time-reversal symmetry, and in the
absence of $\delta$ it also has $M_x$ and $M_y$ mirror and $C_4$
rotational symmetries.  (Strictly speaking, the spatial symmetry
operators only return the system to itself after a sign-flip gauge
change, but this does not affect the symmetry arguments.)  The BBH
model was introduced largely for the purpose of investigating the
consequences of symmetry for the bare model ($\delta=0$). The BBH and
subsequent papers have shown that the presence of $M_x$ and $M_y$
symmetries, or $C_4$ symmetry, constrains the corner charge of a
rectangular flake to be a multiple of $e/4$ quite generally, or of
$e/2$ in some cases~\cite{benalcazar-sa17,benalcazar-prb17,
vanmiert-prb18,benalcazar-prb19,schindler-prresearch19,
watanabe-prb20,kooi-npjqm21},
stimulating interest in the theory of higher-order topological
phases~\cite{parameswaran-p17}. We can understand this in the context
of our \eq{Qcsumd} by noting that $\cP_x\pT=-\cP_y\pR$ and $\cQxy=0$
in a $C_4$-respecting gauge, leaving only the $Q\pTR$ contribution of
\eq{Qf}.  For a general rectangular-lattice system, this must be
either zero or a multiple of $e/4$ (mod $e$), depending on whether any
fractional ionic charges were left over in the corner tile after the
bulk and edge tiling.  (In the context of the BBH model, $C_4$
symmetry implies $\Qc=0$ or $e/2$.)

Here, instead, we are more interested in the case that
spatial symmetries other than inversion are \textit{not} present,
so that the corner charge is not quantized.
Returning to the BBH model, at $\delta=0$ the model has two
gapped phases, a trivial phase with $\Qc=0$
for $|\gamma/\lambda|>1$ and a topological phase with $\Qc=\pm e/2$
for $|\gamma/\lambda|<1$.  The bulk energy gap closes
at the M point in the BZ at the critical $|\gamma/\lambda|=1$.
In what follows a small $\delta$ is applied to break the mirror
and $C_4$ symmetries. Note that we continue to refer to the resulting
systems as being in the ``trivial'' or ``topological'' phase,
even though such a classification is no longer strictly well defined.

\subsubsection{Trivial and topological phases}
\seclab{triv-topo}

In our calculations we set $\lambda=1.0$, and choose $\gamma=1.5$ and
$\gamma=0.5$ to put the system in the trivial and topological phases,
respectively. The resulting energy dispersions, plotted in
\fref{bbh-band-model}(b), consist of two doubly-degenerate bands
separated by finite gaps.  To fix the sign of the corner charge in the
topological phase, BBH weakly broke the quantizing symmetries $M_x$,
$M_y$ and $C_4$ while preserving inversion symmetry by adding a
nonzero $\delta$ term to the
Hamiltonian~\cite{benalcazar-sa17,benalcazar-prb17}. When $\delta$ is
small, $\Qc$ deviates slightly from the quantized value.  The results
reported below are obtained using $\delta=0.001$ for both phases.
Since the model has two occupied bands, we know from our previous
example that the transverse-first nested Wannier construction
cannot be trusted to produce consistent gauges for the two ribbons, so
we focus here on the projection approach from the outset.

\begin{figure}
\centering\includegraphics[width=\columnwidth]{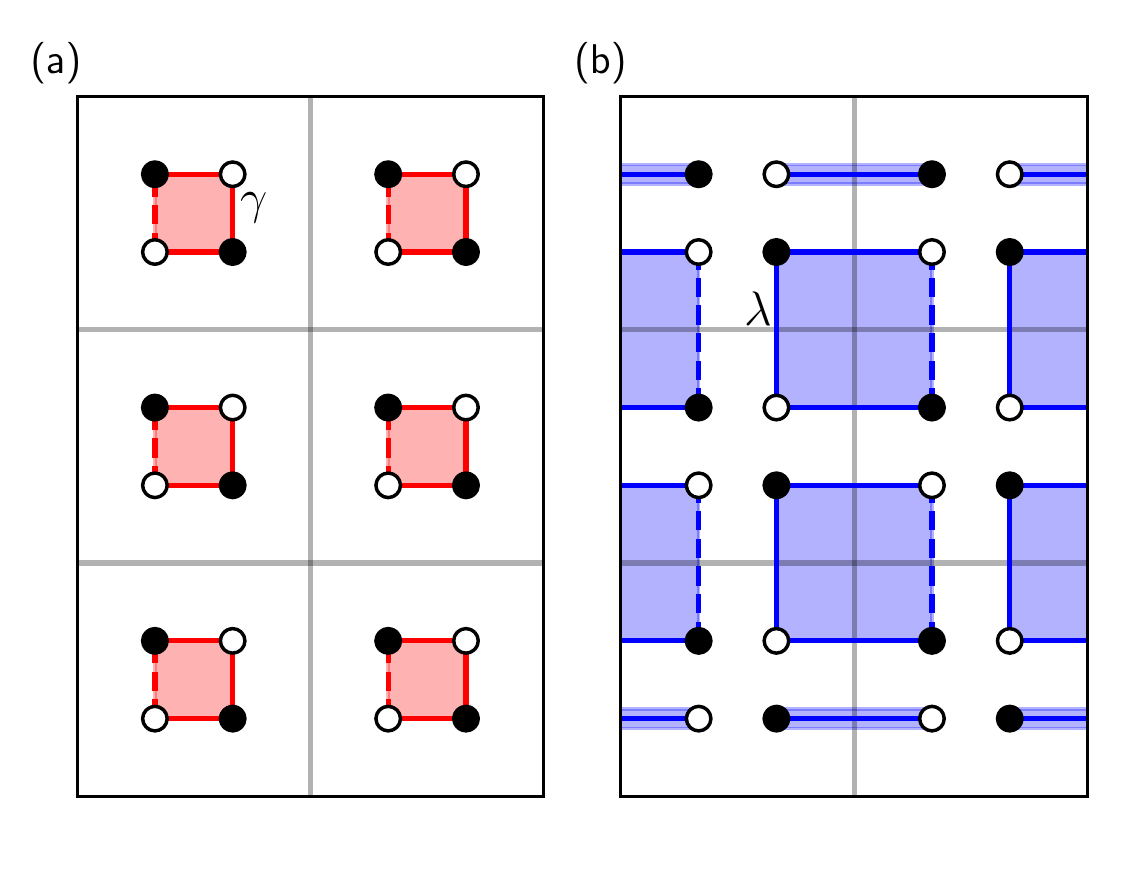}
\caption{Visualization of the isolated tiles whose low-energy
  eigenstates serve as trial functions for constructing Wannier
  functions in ribbons of the Benalcazar-Bernevig-Hughes model.
  The figures represent $y$-finite ribbons three unit cells
  high, while those used in the actual calculations are 40 unit
  cells high.  (a) Tiles used for the trivial phase (red squares).
  (b) Tiles used for the topological phase. The blue squares are
  interior tiles. At the edges and corners, there are ``left-over''
  dimers and isolated atoms, respectively.}
\figlab{bbh-trial-function}
\end{figure}

In view of the qualitative difference between the trivial
and topological phases, we adopt a different choice of bulk tile
for each case.  For the trivial phase we choose the bulk tile to
correspond to the unit cell centered on the small red square in
\fref{bbh-trial-function}(a), with reference locations
$\bm{\tref}_1=\bm{\tref}_2=\0$.  The ionic interior quadrupole in \eq{qxyif}
is thus $q\bxy\pion=0$,
with $q\bxy\pel$ in \eq{qxyef} left to be determined by the anisotropy
of the Wannier charge distribution.
Since we assume the sample has been cut as shown in \fref{bbh-band-model}(a),
there will be no left-over charges in the top-right corner tile,
i.e., $Q\pTR=0$.

By contrast, for the topological phase, the WFs are chosen
to be associated with a large blue square
in \fref{bbh-trial-function}(b), e.g., the one centered
at (1/2,1/2). A choice of tile involving these WFs together with
ions centered around (0,0) would generate
a bulk dipole, which our approach requires us to avoid.  Instead,
we choose the bulk tile as the
unit cell centered on the
large blue square in \fref{bbh-trial-function}(b), with reference
positions $\bm{\tref}_1=\bm{\tref}_2=(1/2,1/2)$ at the center of this
square.\footnote{Note that here, if we had taken a model with ionic
  charge $2e$ at the origin from the outset, we would have needed
  to use the split-basis approach discussed below \eq{dmuI},
  yielding the same pattern of fractional charges.  This is not
  the case for the trivial phase.}
Again the symmetry is such that $q\bxy\pion=0$ in \eq{qxyif}, and
$q\bxy\pel$ in \eq{qxyef} is left to be determined.  Note that there
are now two left-over ionic charges that need to be assigned to each
top tile as shown in \fref{bbh-trial-function}(b), and similarly for
the right edge tiles.  Each corner tile acquires one ionic charge of
$+e/2$, so $Q\pTR$ of the corner tile will be $-e/2$ if there is an
occupied WF in that tile and $+e/2$ otherwise.  From a minimal
knowledge of the model, we can anticipate that a WF will be present in
the top-right tile if and only if $\delta<0$.

\begin{table}
\caption{\label{tab:bbh-wf-pm} Bulk-like Wannier functions in the home
  unit cell of the BBH model in the trivial phase. The Wannier
  functions are constructed using the projection method, choosing as
  trial orbitals $\ket{g_1}$ and $\ket{g_2}$ the lowest-energy
  eigenstates of the isolated red-square tile in
  \fref{bbh-trial-function}(a). $\ket{\phi_{\R j}}$ is the basis
  orbital located at site $\R+\taub_j$, given in reduced coordinates.}
\begin{ruledtabular}
\begin{tabular}{ccccc}
$\R+\taub_j$ & $\ip{\phi_{\R j}}{w_{{\rm int},1}^{({\rm p})}}$ & $\ip{\phi_{\R j}}{g_1}$ & $\ip{\phi_{\R j}}{w_{{\rm int},2}^{({\rm p})}}$ & $\ip{\phi_{\R j}}{g_2}$ \\
\colrule
$(\phm\frac{1}{6},\phm\frac{1}{6})$  &  $\phm0.67899$  &  $\phm0.70694$  &  $0$  &  $0$  \\
$(-\frac{1}{6},-\frac{1}{6})$  &  $0$  &  $0$  &  $\phm0.67899$  &  $\phm0.70694$  \\
$(-\frac{1}{6},\phm\frac{1}{6})$  &  $-0.48037$  &  $-0.50012$  &  $\phm0.48037$  &  $\phm0.50012$  \\
$(\phm\frac{1}{6},-\frac{1}{6})$  &  $-0.48037$  &  $-0.50012$  &  $-0.48037$  &  $-0.50012$  \\
$(\phm\frac{5}{6},\phm\frac{1}{6})$  &  $-0.12012$  &  $0$  &  $-0.03317$  &  $0$  \\
$(\phm\frac{1}{6},\phm\frac{5}{6})$  &  $-0.12012$  &  $0$  &  $\phm0.03317$  &  $0$  \\
$(-\frac{5}{6},-\frac{1}{6})$  &  $\phm0.03317$  &  $0$  &  $-0.12012$  &  $0$  \\
$(-\frac{1}{6},-\frac{5}{6})$  &  $\phm0.03317$  &  $0$  &  $\phm0.12012$  &  $0$  \\
$(\phm\frac{5}{6},-\frac{1}{6})$  &  $\phm0.10844$  &  $0$  &  $\phm0.06157$  &  $0$  \\
$(-\frac{1}{6},\phm\frac{5}{6})$  &  $-0.10844$  &  $0$  &  $\phm0.06157$  &  $0$  \\
$(-\frac{5}{6},\phm\frac{1}{6})$  &  $\phm0.06157$  &  $0$  &  $-0.10844$  &  $0$  \\
$(\phm\frac{1}{6},-\frac{5}{6})$  &  $\phm0.06157$  &  $0$  &  $\phm0.10844$  &  $0$  \\
$(\phm\frac{1}{6},\phm\frac{7}{6})$  &  $-0.05019$  &  $0$  &  $-0.00333$  &  $0$  \\
$(\phm\frac{7}{6},\phm\frac{1}{6})$  &  $-0.05019$  &  $0$  &  $\phm0.00333$  &  $0$  \\
$(-\frac{1}{6},-\frac{7}{6})$  &  $\phm0.00333$  &  $0$  &  $-0.05019$  &  $0$  \\
$(-\frac{7}{6},-\frac{1}{6})$  &  $-0.00333$  &  $0$  &  $-0.05019$  &  $0$  \\
\end{tabular}
\end{ruledtabular}
\end{table}

\begin{table}
\caption{\label{tab:bbh-wf-pm-tp} Same as
    Table~\ref{tab:bbh-wf-pm}, but for the topological phase of the BBH
    model. The trial orbitals $\ket{g_1}$ and $\ket{g_2}$ are now
    chosen as the lowest-energy eigenstates of the isolated blue-square
    tile in \fref{bbh-trial-function}(b).
}
\begin{ruledtabular}
\begin{tabular}{ccccc}
$\R+\taub_j$ & $\ip{\phi_{\R j}}{w_{{\rm int},1}^{({\rm p})}}$ & $\ip{\phi_{\R j}}{g_1}$ & $\ip{\phi_{\R j}}{w_{{\rm int},2}^{({\rm p})}}$ & $\ip{\phi_{\R j}}{g_2}$ \\
\colrule
$(\phm\frac{1}{6},\phm\frac{1}{6})$  &  $\phm0.69081$  &  $\phm0.70686$  &  $0$  &  $0$  \\
$(\phm\frac{5}{6},\phm\frac{5}{6})$  &  $0$  &  $0$  &  $\phm0.69081$  &  $\phm0.70686$  \\
$(\phm\frac{5}{6},\phm\frac{1}{6})$  &  $-0.48885$  &  $-0.50018$  &  $\phm0.48885$  &  $\phm0.50018$  \\
$(\phm\frac{1}{6},\phm\frac{5}{6})$  &  $-0.48885$  &  $-0.50018$  &  $-0.48885$  &  $-0.50018$  \\
$(-\frac{1}{6},\phm\frac{1}{6})$  &  $-0.09152$  &  $0$  &  $-0.02753$  &  $0$  \\
$(\phm\frac{1}{6},-\frac{1}{6})$  &  $-0.09152$  &  $0$  &  $\phm0.02753$  &  $0$  \\
$(\phm\frac{7}{6},\phm\frac{5}{6})$  &  $\phm0.02753$  &  $0$  &  $-0.09152$  &  $0$  \\
$(\phm\frac{5}{6},\phm\frac{7}{6})$  &  $\phm0.02753$  &  $0$  &  $\phm0.09152$  &  $0$  \\
$(\phm\frac{5}{6},-\frac{1}{6})$  &  $-0.08424$  &  $0$  &  $\phm0.04535$  &  $0$  \\
$(-\frac{1}{6},\phm\frac{5}{6})$  &  $\phm0.08424$  &  $0$  &  $\phm0.04535$  &  $0$  \\
$(\phm\frac{1}{6},\phm\frac{7}{6})$  &  $\phm0.04535$  &  $0$  &  $\phm0.08424$  &  $0$  \\
$(\phm\frac{7}{6},\phm\frac{1}{6})$  &  $\phm0.04535$  &  $0$  &  $-0.08424$  &  $0$  \\
$(-\frac{5}{6},\phm\frac{1}{6})$  &  $-0.04049$  &  $0$  &  $\phm0.00160$  &  $0$  \\
$(\phm\frac{1}{6},-\frac{5}{6})$  &  $-0.04049$  &  $0$  &  $-0.00160$  &  $0$  \\
$(\phm\frac{11}{6},\phm\frac{5}{6})$  &  $-0.00160$  &  $0$  &  $-0.04049$  &  $0$  \\
$(\phm\frac{5}{6},\phm\frac{11}{6})$  &  $\phm0.00160$  &  $0$  &  $-0.04049$  &  $0$  \\
\end{tabular}
\end{ruledtabular}
\end{table}

To obtain gauge-consistent values for $\cP_x\pT$, $\cP_y\pR$, and
$\cQxy\pI$ via projection, we begin by considering a $y$-finite ribbon
40 unit cells high, with simple periodic boundary conditions along
$x$.  The trial functions are chosen as the low-energy eigenstates of
the isolated tiles obtained by removing the weaker of the two
hoppings.  For the trivial phase, we take as trial functions the two
lowest-energy eigenstates of the isolated small red square
in \fref{bbh-trial-function},
replicated 40 times to cover the entire ribbon.  For the topological
phase the WF centers shift to the large blue
squares~\cite{song-prl17,ezawa-prb18,khalaf-2019arxiv}, so we take
their isolated eigenstates as our projection functions, replicated 39
times.  We also include two edge tiles, one at the top and one at
the bottom of the ribbon, each consisting of a single dimer with its
single low-energy eigenstate.  Taken together, these states comprise
our trial functions for the ribbon in the topological phase.  We do
the same for $x$-finite ribbons, and we confirm that within each
phase, the deep interior WFs are identical for $x$- and $y$-finite
ribbons.  The site amplitudes of the resulting WFs are given in
Table~\ref{tab:bbh-wf-pm} for the trivial phase, and in
Table~\ref{tab:bbh-wf-pm-tp} for the topological phase, together with
the trial functions for comparison.

From the consistent sets of WFs obtained for the two ribbons, we
calculate edge polarizations and interior quadrupoles in the usual
manner. To accommodate the left-over dimer WFs in the outermost layers
in the topological phase, the edge polarizations are evaluated from
edge tiles containing an odd number of WFs, while in the trivial phase
that number is even. The values of $\cP_x\pT$, $\cP_y\pR$, and
$\cQxy\pI$ are listed in Table~\ref{tab:bbh-result}.
These are all very small, of order $10^{-5}\,e$ and
$10^{-4}\,e$ in the trivial and topological phases respectively, as a
consequence of the small $\delta$. The fourth contribution $Q\pTR$
vanishes in the trivial phase and is $e/2$ in the topological phase.
Summing all four contributions, we find excellent agreement with the
directly calculated macroscopic corner charge in both phases.
Thus, in both cases, the
small deviation from the quantized $\Qc$ value caused by the staggered
on-site potential is precisely reproduced by the ribbon
calculations.

\begin{table}
\caption{\label{tab:bbh-result} 
  Individual contributions and total predicted macroscopic
  corner charge $\Qc$ in \eq{Qcsumd}, compared with a direct calculation,
  for the trivial and topological phases of the BBH model as
  depicted in  \fref{bbh-band-model}(b). In both cases, the
  symmetries that quantize the corner charge are weakly broken
  by a staggered on-site potential (see main text).
  The values of $\cP_x\pT$, $\cP_y\pR$, and $\cQxy\pI$ are
  obtained from ribbon calculations, while $Q\pTR$ is
  inferred mod $e$ from the tiling procedure.
  The last line reports the bare corner charge computed by summing over
  the top-right quadrant.
}
\begin{ruledtabular}
\begin{tabular}{lcc}
& Trivial  & Topological \\
\colrule
$\cP_x\pT$  & $0.854\times10^{-5}$ & $-44.077\times10^{-5}$   \\
$\cP_y\pR$  & $0.854\times10^{-5}$ & $-44.077\times10^{-5}$   \\
$\cQxy\pI$ & $4.517\times10^{-5}$ & $\phm18.412\times10^{-5}$  \\
$Q\pTR$ & 0 & $0.5$ \\
\colrule
$\Qc$ (predicted) & $6.225\times10^{-5}$ & $0.5-69.743\times10^{-5}$ \\
$\Qc$ (direct) & $6.225\times10^{-5}$ & $0.5-69.743\times10^{-5}$ \\
$\Qc^{\rm bare}$ & $1.602\times10^{-5}$ & $0.5-84.817\times10^{-5}$
\end{tabular}
\end{ruledtabular}
\end{table}

\subsubsection{Corner charge pumping cycle}
\seclab{pump}

In this section, we carry out calculations of the interior
quadrupole and edge polarizations, and compare the predicted
corner charge with the directly calculated one, for the same
adiabatic cycle
\beq
(\delta,\lambda,\gamma) = \Bigg\{
\begin{matrix}
        (\cos(t),\sin(t),0) \quad 0<t\leqslant\pi
        \\ (\cos(t),0,\lvert \sin(t) \rvert) \quad \pi<t\leqslant2\pi
\end{matrix} 
\eqlab{adiabatic}
\eeq
considered previously by BBH~\cite{benalcazar-prb17,benalcazar-sa17}.
This cycle is somewhat artificial, in that one or the other of the
hoppings $\gamma$ or $\lambda$ is always zero. However, to make contact
with previous literature, we apply our method to the
same system here.

At $t=0$ the system starts in a state in which the sites are
completely decoupled, with only black sites in
\fref{bbh-trial-function} occupied as a result of the positive
$\delta$.  In the interval $0<t<\pi$, a set of positive $\lambda$
hoppings are first turned on and then turned off on the edges of the
large blue squares in \fref{bbh-trial-function}.  In this interval,
the system takes the form of a molecular crystal with ``molecules''
centered on the large blue squares.  At $t=\pi/2$ where $\delta$
vanishes, the symmetry suffices to define the topological index, and
the system is in the nontrivial phase.  Once $t$ passes $\pi/2$ the
sign of $\delta$ is reversed, so that at $t=\pi$ we again reach a
state of completely decoupled sites, but now with only the open-circle
sites occupied.  The second half of the loop is similar, except that
now the $\gamma$ hoppings are progressively turned on and off, so that
the system is molecular once more, but centered on the small red
squares. The topology is again defined at $t=3\pi/2$, now being
trivial, and the system returns to its starting point at $t=2\pi$.

We use two different sets of trial functions for
the Wannier projection during the first and second halves
of the cycle.
For $t\in[0,\pi]$ we adopt the trial functions of the topological
state, while for $t\in[\pi,2\pi]$ we choose those of the trivial
state, as described in the previous subsection and detailed in Tables
VI and V respectively.  We thus have a gauge discontinuity at $t=\pi$
and again at $t=2\pi$.  For a mesh of $t$ values, we compute
$\cQxy\pI$, $\cP_x\pT$, and $\cP_y\pR$, and compare the prediction of
\eq{Qcsumd} with the directly computed macroscopic corner charge of a
large but finite flake.  The results are presented in
\fref{bbh-adiabatic}.  Since the corner charge is predicted only mod
$e$, we plot several branches corresponding to the periodicity of $e$
along the vertical axis as blue dots, and the directly calculated
corner charges are the red circles.

\begin{figure}
\centering\includegraphics[width=\columnwidth]{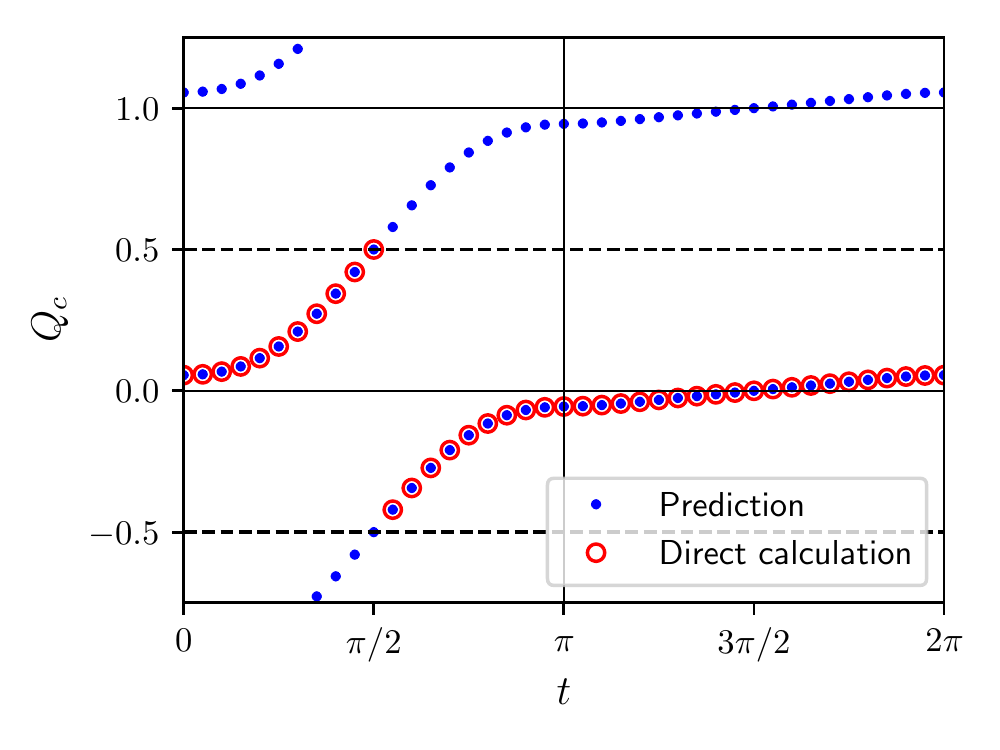}
\caption{Evolution of the corner charge $\Qc$ (in units of $e$) during
  the adiabatic pumping cycle described by \eq{adiabatic}. The ribbon
  calculations only predict $\Qc$ modulo $e$, so three branches are
  plotted vs.\ $t$ as the blue dots. The evolution of the actual
  corner charge of a finite flake is indicated by the red circles.}
\figlab{bbh-adiabatic}
\end{figure}

We confirm that $\Qc=e/2$ and zero (mod $e$)
at $\pi/2$ and $3\pi/2$ respectively,
where the topology is sharply defined.  However, we find that most
of the pumping of the corner charge occurs in the first half of the
cycle. That is, $\Qc$ grows from $e/18$ to $17e/18$ in this interval,
for an increase of $8e/9$, while the growth in the second half of
the cycle is only by the remaining amount $e/9$.

Note that the gauge discontinuities at $t=\pi$ and $2\pi$
introduce no discontinuities in the
predicted value of $\Qc$.  However, there are
discontinuities in
the individual values of $\cQxy\pI$, $\cP_x\pT$, and $\cP_y\pR$.  In
the first half of the cycle, $\cQxy\pI$ comes from the larger
blue-square tile and changes from $2e/9$ to $-2e/9$, while in the
second half $\cQxy\pI$ comes from the twice-smaller
red-square tile and grows from $-e/18$ to $e/18$.
In the first half-cycle, $\cP_x\pT=\cP_y\pR$ each increase from
$-e/3$ to $e/3$, while in the second half $\cP_x\pT$ and $\cP_y\pR$
are identically zero.  Finally, our tiling is such that
$Q\pTR=e/2$ (mod $e$) in the first half-cycle, and zero (mod $e$)
in the second half.  Adding the various
contributions according to \eq{Qcsumd}, we find that the total
$\Qc$ evolves as described in the previous paragraph.

Without a knowledge of the population of WFs in the corner tile,
we can only make predictions ``mod $e$'' as done above.
In particular, we cannot predict precisely when the corner charge
will make the discontinuous jump needed to allow it to return
to its initial state at the end of the pumping cycle.  However,
by inspecting the Hamiltonian,
we can anticipate that a WF will be present in the
top-right tile in the interval $\pi/2<t<\pi$,
when the open circle
at top right in \fref{bbh-trial-function}(b) has negative energy,
but not otherwise.  Making use of this additional information
about $Q\pTR$, we expect the discontinuity in the macroscopic
corner charge to occur at $t=\pi/2$.  We then correctly predict
not only the value mod $e$, but also the correct
branch choice, of $\Qc$ over the entire cycle.

A comparison of our \fref{bbh-adiabatic} with Fig.~37 of
Ref.~[\onlinecite{benalcazar-prb17}], which also compares predictions
from ribbons with a computed corner charge, shows important
differences.  In their case, all the change in the corner charge
occurs in the first half-cycle, when it evolves from 0 to $e$, and
there is no change in the second half-cycle.  While the computed
corner charges agree with the predictions in their theory, as they do
in ours, it is important to keep in mind that the two approaches
differ in crucial ways.  (i) In Ref.~[\onlinecite{benalcazar-prb17}],
BBH do not compute the {\it macroscopic} corner charge defined by
\eq{qcbar}; instead, they compute the total charge of the upper-right
quadrant, that is, the {\it bare} corner charge of \eq{qcnom}.
In fact, since they did not
specify the positions of the orbitals, the
macroscopic corner charge is ill-determined in their case.
For the trivial and topological cases discussed in \sref{triv-topo},
we obtain the values of $Q_c^{\rm bare}$ presented in the last row
of Table~\ref{tab:bbh-result}, which are clearly very different
from the macroscopic corner charges.%
\footnote{If all sites are located precisely at the origin in
  the middle of the unit cell,
  the bare and macroscopic corner charges become equal.
  This follows because $W(x,y)$ in \eq{qcW} is identical for all
  electronic and ionic charges in the cell, and the total
  charge of the cell vanishes both for deep interior cells
  and for skin cells far from the corners.  Thus,
  \eq{qcW} is equivalent to integrating the charge density
  over a quadrant.}
(ii) Their
edge polarizations $p^{\rm edge}$ are not defined in the same way as
ours.  For the specified cycle, their $p^{\rm edge}$ is defined in
such a way that $dp^{\rm edge}/dt$ corresponds to the flow of current
into a quadrant, while our $d\cP/dt$ corresponds to the polarization
current associated with the changing dipole moments of the edge tiles
in the skin region.
(iii) In our theory, in order to correctly predict the macroscopic
corner charge, we also insist that bulk quadrupole and surface
dipole contributions are computed in a common Wannier gauge.
As a result of these differences, each
theory obtains internally consistent results, although we argue that
ours is more physical in that it predicts a macroscopically observable
corner charge.

\section{Gauge-consistent nested Wannier construction}
\seclab{gcnw}

In the previous section, we demonstrated that a naive
application of the hybrid Wannier approach, in which the
transverse-first nested Wannier construction is applied
to ribbons of both orientations, is not gauge-consistent,
whereas an alternative projection construction does result in
a consistent gauge.  Here, we demonstrate a second successful
method for generating a consistent gauge, this time without the
need for providing trial functions.  We do this using the nested
Wannier constructions described in \srefs{nw-trans}{nw-long}, but
now insuring that the two localization steps are executed
in the same order for both the $x$-finite and $y$-finite ribbons.
In other words, one should apply the transverse
construction of \sref{nw-trans} to one ribbon, and the longitudinal
construction of \sref{nw-long} to the other.

Let us apply this procedure to the four-band model of
\fref{4-band-model}, for which we obtained inconsistent gauges in
\sref{4-band} by applying the transverse construction to both
ribbons. We choose to localize first along $y$ and then along
$x$. Thus we apply the same transverse construction as before to the
$y$-finite ribbon, and apply the longitudinal construction to the
$x$-finite ribbon.  We find that deep inside the two ribbons the
resulting WFs are identical: within numerical accuracy, their site
amplitudes are the same and the quantum distance between them
vanishes. We then repeat the entire procedure but localizing first
along $x$ and then along $y$, and again we arrive at the same interior
gauge for both ribbons (but different from the previous one).

Table~\ref{tab:4-band-hwf-result} shows the individual contributions
and total predicted corner charge in the two nested Wannier
gauges. The predicted corner charges are the same in both, and they
agree perfectly with the actual corner charge of a finite flake.  Note
that while the edge polarizations are different between those two
gauges, the interior quadrupoles are identical. The reason is that
$\cQxy\pI$ is a symmetric tensor, and hence it remains unchanged upon
reversing the order of the $x$ and $y$ localization steps.

We have also tested this gauge-consistent nested Wannier
approach for the BBH model~\cite{benalcazar-prb17,benalcazar-sa17},
and we again find that the corner charge is correctly predicted.
The implementation is straightforward following the example of
the four-band-model discussed above.

Before concluding the discussion of this method, we note
that it is possible to bypass the second step of the
longitudinal-first construction. Briefly, again working in the $y$-first
context, we carry out only the first step of the $y$-first
construction for the $x$-finite ribbon.  We identify the
total charge $\rho_{\ell_y}(\r)$ of the ions and
WFs associated with any one of the single-cell-high layers
$\ell_y$, and compute its $y$-dipole density
$d^{(y)}(x)=\int y\,\rho_{\ell_y}(x,y)\,dy$.
This quantity is independent of $\ell_y$, and letting
$\bar{d}^{(y)}(x)$ be its window average in the $x$ direction,
we note that $\bar{d}^{(y)}(x)$ vanishes except near the edges of
the ribbon, and its integral over the right skin region gives
$\cQxy\pI+\cP_y\pR$.  Adding this to the $\cP_x\pT$ obtained
from the transverse-first nested Wannier construction for
the $y$-finite ribbon then gives the correct corner charge as
before. Nevertheless, we recommend applying the two-step nested
procedure to both ribbons, as this increases the reliability of
the method by allowing a cross-check on the equivalence of the
two sets of WFs.

\begin{table}
\caption{\label{tab:4-band-hwf-result} Individual contributions and total
        predicted macroscopic corner charge $\Qc$ in \eq{Qcsumd},
        compared with a direct calculation, for the four-band model of
        \fref{4-band-model}. The ribbon calculations were performed
        using a gauge-consistent nested Wannier construction where
        we first Wannierize along $y$ and then along $x$ (middle
        column), or vice-versa (right-column).}
\begin{ruledtabular}
\begin{tabular}{lcc}
&  Wannierize $y$ then $x$ & Wannierize $x$ then $y$ \\
&  $(10^{-2}~e)$ & $(10^{-2}~e)$ \\
\colrule
$\cP_x\pT$ & $\phm0.300250$ & $\phm0.296029$   \\
$\cP_y\pR$ & $\phm0.472198$ & $\phm0.476420$   \\
$\cQxy\pI$ & $-3.756016$ & $-3.756016$   \\
\colrule
$\Qc$ (predicted) & $-2.983567$ & $-2.983567$  \\
$\Qc$ (direct) & $-2.983567$ & $-2.983567$
\end{tabular}
\end{ruledtabular}
\end{table}

\section{Discussion}
\seclab{discuss}

Several generalizations of our work remain to be developed.
Our current formulation is trivially extended to the case of
broken time-reversal symmetry, and the presence of spinor electrons
entails no special difficulty.
The case of nonrectangular crystals and corner
angles other than 90$^\circ$ can be treated following
the methods of Ref.~[\onlinecite{trifunovic-prresearch20}].
By contrast, generalizations to topological systems, such as
2D Chern insulators or $Z_2$-odd quantum spin Hall insulators,
do not look straightforward. In these cases, metallic edge
states are topologically protected, interfering with any
natural definition of edge polarization.
Finally, while we have focused here on the case of low-symmetry
systems such that the corner charge is not quantized,
further exploration of the connections to the theory of
higher-order topological insulators in higher-symmetry
systems is desirable.

Generalizations to higher dimensions are easily anticipated.
The line of intersection of two surface facets of a 3D crystal,
generally known as a ``hinge,'' carries a linear charge density
that can be computed via an elementary extension of the present
methods, either by Wannierizing in all three dimensions, or by
Wannierizing in 2D at each $k_\parallel$ (wavevector along the hinge)
and averaging over $k_\parallel$.  The prediction of the
corner charge in 3D, while perhaps more difficult in practice,
should follow the same principles outlined here.
That is, one would need to compute the octupoles of interior
bulk tiles far from any surfaces,
the quadrupoles of surface tiles far from any hinges,
and the dipoles of hinge tiles.  While these will not be
individually gauge-invariant, their sum will be, allowing for
a prediction of the corner charge mod $e$.  So, for example,
a calculation of three rectangular rod geometries, one each
extending along $\xhat$, $\yhat$, and $\zhat$, should provide
all the needed information.

Throughout this work we have assumed the presence of bulk
inversion symmetry so that the bulk cell can be chosen
to be free of an electric dipole moment.  However,
other symmetries can also force a nonpolar point group.   In
2D these would be the $C_6$, $C_4$, and $C_3$ rotations
($C_2$ is equivalent to inversion in 2D). All of these $C_n$ symmetries
force $q_{xy}\pI$ to vanish, and result in quantized corner charges for
a crystallite in the shape of a regular $n$-gon.  However, there could
be cases of inequivalent edges meeting, as for example a
90$^\circ$ corner of a material with bulk $C_3$ symmetry.
In such cases the adjoining edges are inequivalent and could
result in a generic corner charge.
More opportunities arise for nonpolar but noncentrosymmetric
point groups in 3D.  It should be straightforward to generalize
our theory to such cases.

We end this section with a discussion of connections to the
theory or orbital magnetization, which we already briefly invoked
to argue that surface polarization is not a physical observable.
We argued that if it were, its time derivative ought to correspond
to a physical flow of current at the edge of the 2D sample.
However, for a time-reversal broken system with a nonzero orbital
magnetization, a steady current circulates around the edges of
the sample, which is inconsistent with a uniquely defined edge
polarization.  By contrast, it is clear that the edge current
\textit{is} a physical observable; it can be evaluated as an
expectation value of a Hermitian operator in the usual way,
and is fully gauge-invariant.

There is a strong formal similarity between the theory presented here
and that developed by
Thonhauser \textit{et al.}~\cite{thonhauser-prl05} and
Ceresoli \textit{et al.}~\cite{ceresoli-prb06} to derive the modern-theory
expression
for orbital magnetization using the Wannier representation.  In fact,
that work made use of an identical decomposition of the Wannier
functions of a large but finite flake into those associated with
interior and skin regions, and identified two contributions to the
orbital magnetization.  One, denoted as the ``local circulation,'' was
identified with the internal circulation of charge in a deep-interior
WF.  The second, labeled ``itinerant circulation,'' arises from edge
currents defined as the expectation value of the current operator
traced over WFs in the skin region.  The current of this type on the
right-hand edge, labeled as $I_y$ in
Ref.~[\onlinecite{thonhauser-prl05}] and denoted as $I_y\pR$
henceforth, is just the time derivative of the edge polarization
$\cP_y\pR$ defined here.  Indeed the expression for $I_y\pR$ in
Eq.~(9) of Ref.~[\onlinecite{thonhauser-prl05}] takes the form of a
sum of contributions from hoppings that cross the boundary between the
interior and skin regions, just as our expression in \eq{Pyvar} for
the change in $\cP_y\pR$ under a gauge change depends on lattice
vectors $\R'$ crossing that same boundary.

This is no accident.  Since we are in the ground state, the
unitary time-evolution operator $e^{-iHt/\hbar}$ does not change the
occupied subspace, but it does modify the gauge by multiplying each
energy eigenstate by a phase factor $e^{-iEt/\hbar}$.  An
infinitesimal time step $\delta t$ corresponds to an
infinitesimal unitary transformation in which
the deep interior WFs change by
$\delta\ket{\0 m}=\sum_{\R'n}\eps_{\R',nm}\ket{\R' n}$, using a
notation consistent with \eq{gc-eps}, with
\beq
\eps_{\R',nm}=-i\frac{\delta t}{\hbar}\me{\R'n}{H}{\0 m} \,.
\eeq
Substituting into \eq{Pyvar} and using \eq{XYdef}, the
upward-flowing current $I_y\pR=\delta \cP_y\pR/\delta t$ on the
right edge of the sample is
\beq
I_y\pR=\frac{2e}{ab}\frac{1}{\hbar}
  \sum_{R'_x>0}\sum_{nm} R'_x \me{\0 m}{y}{\R'n} \me{\R'n}{H}{\0 m} \,.
\eqlab{Iy}
\eeq
In other words, time evolution within the occupied subspace
generates a gauge evolution, and the changing gauge drives
a displacement of WF centers in the skin region that corresponds
precisely to the itinerant edge current $I_y\pR$.
\Eq{Iy} reproduces the expressions derived in
Refs.~\cite{thonhauser-prl05} and \cite{ceresoli-prb06} for the
single-band and multiband cases respectively.
The (counterclockwise) itinerant-circulation contribution to the
orbital magnetization is given by the \textit{average} of $I_y\pR$
on the right edge and $-I_x\pT$ on the top edge, while instead
the \textit{difference} between $I_y\pR$ and $-I_x\pT$ (that is,
$I_y\pR+I_x\pT$) corresponds to a skin contribution to the time
rate of change of the top-right corner charge. The latter is in fact
independent of time, so this must be exactly canceled by a contribution
from the time dependence of the interior-tile Wannier quadrupole,
which is more closely related to the local circulation in the
orbital magnetization theory.

These relationships indicate a deep formal connection between
the theory of orbital magnetization and that of edge polarizations
and corner charges presented here.

\section{Summary}
\seclab{summary}

In summary, we have considered the case of a 2D centrosymmetric
insulator in which the corner charges are not quantized by additional
symmetries. Decomposing the large but finite flake into bulk, skin,
and corner regions, and introducing a tiling in this context, we have
shown that the corner charge can be written as a sum of a quadrupole
contribution associated with the bulk tiles,
and two dipole contributions 
associated with the two edges that meet at the corner.
Having introduced a Wannier representation to
attach electron charges to these tiles, we demonstrated that the bulk
quadrupole and two edge dipole contributions are not individually
gauge-invariant, although their sum is. As a consequence, we argue
that it is crucially important to adopt a common gauge for the
computation of all of these quantities in the two ribbon geometries.

To verify the correctness of our approach, we have
tested it via calculations on three different tight-binding
models.  We have demonstrated two different methods for arriving
at a consistent gauge for ribbons of both orientations,
one based on projection from trial functions and another based on
a consistently applied nested Wannier construction.
We emphasize that the macroscopically observable
corner charge has to be computed by an appropriate coarse-graining
procedure, and not simply by counting charges in a quadrant of the
sample.  Having taken all these constraints into account, we
have demonstrated that the corner charge can indeed be computed
modulo $e$, to numerical accuracy, from calculations on two ribbon
geometries alone.
We are hopeful that our work paves the way toward
the emergence of a deeper and more general understanding of the
intimate connections between bulk and surface properties of
crystalline materials.

\section*{Acknowledgments}

Work by S.R.~and D.V.~was supported by NSF Grant DMR-1954856.
Work by I.S. was supported by Grant No.~FIS2016-77188-P from the
Spanish Ministerio de Econom\'ia y Competitividad.

\bibliography{pap}

\begin{thebibliography}{32}%
\makeatletter
\providecommand \@ifxundefined [1]{%
 \@ifx{#1\undefined}
}%
\providecommand \@ifnum [1]{%
 \ifnum #1\expandafter \@firstoftwo
 \else \expandafter \@secondoftwo
 \fi
}%
\providecommand \@ifx [1]{%
 \ifx #1\expandafter \@firstoftwo
 \else \expandafter \@secondoftwo
 \fi
}%
\providecommand \natexlab [1]{#1}%
\providecommand \enquote  [1]{``#1''}%
\providecommand \bibnamefont  [1]{#1}%
\providecommand \bibfnamefont [1]{#1}%
\providecommand \citenamefont [1]{#1}%
\providecommand \href@noop [0]{\@secondoftwo}%
\providecommand \href [0]{\begingroup \@sanitize@url \@href}%
\providecommand \@href[1]{\@@startlink{#1}\@@href}%
\providecommand \@@href[1]{\endgroup#1\@@endlink}%
\providecommand \@sanitize@url [0]{\catcode `\\12\catcode `\$12\catcode
  `\&12\catcode `\#12\catcode `\^12\catcode `\_12\catcode `\%12\relax}%
\providecommand \@@startlink[1]{}%
\providecommand \@@endlink[0]{}%
\providecommand \url  [0]{\begingroup\@sanitize@url \@url }%
\providecommand \@url [1]{\endgroup\@href {#1}{\urlprefix }}%
\providecommand \urlprefix  [0]{URL }%
\providecommand \Eprint [0]{\href }%
\providecommand \doibase [0]{http://dx.doi.org/}%
\providecommand \selectlanguage [0]{\@gobble}%
\providecommand \bibinfo  [0]{\@secondoftwo}%
\providecommand \bibfield  [0]{\@secondoftwo}%
\providecommand \translation [1]{[#1]}%
\providecommand \BibitemOpen [0]{}%
\providecommand \bibitemStop [0]{}%
\providecommand \bibitemNoStop [0]{.\EOS\space}%
\providecommand \EOS [0]{\spacefactor3000\relax}%
\providecommand \BibitemShut  [1]{\csname bibitem#1\endcsname}%
\let\auto@bib@innerbib\@empty
\bibitem [{\citenamefont {King-Smith}\ and\ \citenamefont
  {Vanderbilt}(1993)}]{king-smith-prb93}%
  \BibitemOpen
  \bibfield  {author} {\bibinfo {author} {\bibfnamefont {R.~D.}\ \bibnamefont
  {King-Smith}}\ and\ \bibinfo {author} {\bibfnamefont {David}\ \bibnamefont
  {Vanderbilt}},\ }\bibfield  {title} {\enquote {\bibinfo {title} {Theory of
  polarization of crystalline solids},}\ }\href {\doibase
  10.1103/PhysRevB.47.1651} {\bibfield  {journal} {\bibinfo  {journal} {Phys.
  Rev. B}\ }\textbf {\bibinfo {volume} {47}},\ \bibinfo {pages} {1651--1654}
  (\bibinfo {year} {1993})}\BibitemShut {NoStop}%
\bibitem [{\citenamefont {Resta}(1994)}]{resta-rmp94}%
  \BibitemOpen
  \bibfield  {author} {\bibinfo {author} {\bibfnamefont {Raffaele}\
  \bibnamefont {Resta}},\ }\bibfield  {title} {\enquote {\bibinfo {title}
  {Macroscopic polarization in crystalline dielectrics: the geometric phase
  approach},}\ }\href {\doibase 10.1103/RevModPhys.66.899} {\bibfield
  {journal} {\bibinfo  {journal} {Rev. Mod. Phys.}\ }\textbf {\bibinfo {volume}
  {66}},\ \bibinfo {pages} {899--915} (\bibinfo {year} {1994})}\BibitemShut
  {NoStop}%
\bibitem [{\citenamefont {Vanderbilt}(2018)}]{vanderbilt-book18}%
  \BibitemOpen
  \bibfield  {author} {\bibinfo {author} {\bibfnamefont {D.}~\bibnamefont
  {Vanderbilt}},\ }\href {\doibase 10.1017/9781316662205} {\emph {\bibinfo
  {title} {Berry Phases in Electronic Structure Theory}}}\ (\bibinfo
  {publisher} {Cambridge University Press},\ \bibinfo {year}
  {2018})\BibitemShut {NoStop}%
\bibitem [{\citenamefont {Benalcazar}\ \emph
  {et~al.}(2017{\natexlab{a}})\citenamefont {Benalcazar}, \citenamefont
  {Bernevig},\ and\ \citenamefont {Hughes}}]{benalcazar-sa17}%
  \BibitemOpen
  \bibfield  {author} {\bibinfo {author} {\bibfnamefont {W.~A.}\ \bibnamefont
  {Benalcazar}}, \bibinfo {author} {\bibfnamefont {B.~A.}\ \bibnamefont
  {Bernevig}}, \ and\ \bibinfo {author} {\bibfnamefont {T.~L.}\ \bibnamefont
  {Hughes}},\ }\bibfield  {title} {\enquote {\bibinfo {title} {Quantized
  electric multipole insulators},}\ }\href {\doibase 10.1126/science.aah6442}
  {\bibfield  {journal} {\bibinfo  {journal} {Science}\ }\textbf {\bibinfo
  {volume} {357}},\ \bibinfo {pages} {61--66} (\bibinfo {year}
  {2017}{\natexlab{a}})}\BibitemShut {NoStop}%
\bibitem [{\citenamefont {Benalcazar}\ \emph
  {et~al.}(2017{\natexlab{b}})\citenamefont {Benalcazar}, \citenamefont
  {Bernevig},\ and\ \citenamefont {Hughes}}]{benalcazar-prb17}%
  \BibitemOpen
  \bibfield  {author} {\bibinfo {author} {\bibfnamefont {W.~A.}\ \bibnamefont
  {Benalcazar}}, \bibinfo {author} {\bibfnamefont {B.~A.}\ \bibnamefont
  {Bernevig}}, \ and\ \bibinfo {author} {\bibfnamefont {T.~L.}\ \bibnamefont
  {Hughes}},\ }\bibfield  {title} {\enquote {\bibinfo {title} {Electric
  multipole moments, topological multipole moment pumping, and chiral hinge
  states in crystalline insulators},}\ }\href {\doibase
  10.1103/PhysRevB.96.245115} {\bibfield  {journal} {\bibinfo  {journal} {Phys.
  Rev. B}\ }\textbf {\bibinfo {volume} {96}},\ \bibinfo {pages} {245115}
  (\bibinfo {year} {2017}{\natexlab{b}})}\BibitemShut {NoStop}%
\bibitem [{\citenamefont {Parameswaran}\ and\ \citenamefont
  {Wan}(2017)}]{parameswaran-p17}%
  \BibitemOpen
  \bibfield  {author} {\bibinfo {author} {\bibfnamefont {Siddharth~A}\
  \bibnamefont {Parameswaran}}\ and\ \bibinfo {author} {\bibfnamefont {Yuan}\
  \bibnamefont {Wan}},\ }\bibfield  {title} {\enquote {\bibinfo {title}
  {Topological insulators turn a corner},}\ }\href@noop {} {\bibfield
  {journal} {\bibinfo  {journal} {Physics}\ }\textbf {\bibinfo {volume} {10}},\
  \bibinfo {pages} {132} (\bibinfo {year} {2017})}\BibitemShut {NoStop}%
\bibitem [{\citenamefont {Song}\ \emph {et~al.}(2017)\citenamefont {Song},
  \citenamefont {Fang},\ and\ \citenamefont {Fang}}]{song-prl17}%
  \BibitemOpen
  \bibfield  {author} {\bibinfo {author} {\bibfnamefont {Zhida}\ \bibnamefont
  {Song}}, \bibinfo {author} {\bibfnamefont {Zhong}\ \bibnamefont {Fang}}, \
  and\ \bibinfo {author} {\bibfnamefont {Chen}\ \bibnamefont {Fang}},\
  }\bibfield  {title} {\enquote {\bibinfo {title}
  {$(d\ensuremath{-}2)$-dimensional edge states of rotation symmetry protected
  topological states},}\ }\href {\doibase 10.1103/PhysRevLett.119.246402}
  {\bibfield  {journal} {\bibinfo  {journal} {Phys. Rev. Lett.}\ }\textbf
  {\bibinfo {volume} {119}},\ \bibinfo {pages} {246402} (\bibinfo {year}
  {2017})}\BibitemShut {NoStop}%
\bibitem [{\citenamefont {van Miert}\ and\ \citenamefont
  {Ortix}(2018)}]{vanmiert-prb18}%
  \BibitemOpen
  \bibfield  {author} {\bibinfo {author} {\bibfnamefont {Guido}\ \bibnamefont
  {van Miert}}\ and\ \bibinfo {author} {\bibfnamefont {Carmine}\ \bibnamefont
  {Ortix}},\ }\bibfield  {title} {\enquote {\bibinfo {title} {Higher-order
  topological insulators protected by inversion and rotoinversion
  symmetries},}\ }\href {\doibase 10.1103/PhysRevB.98.081110} {\bibfield
  {journal} {\bibinfo  {journal} {Phys. Rev. B}\ }\textbf {\bibinfo {volume}
  {98}},\ \bibinfo {pages} {081110} (\bibinfo {year} {2018})}\BibitemShut
  {NoStop}%
\bibitem [{\citenamefont {Ezawa}(2018)}]{ezawa-prb18}%
  \BibitemOpen
  \bibfield  {author} {\bibinfo {author} {\bibfnamefont {Motohiko}\
  \bibnamefont {Ezawa}},\ }\bibfield  {title} {\enquote {\bibinfo {title}
  {Minimal models for wannier-type higher-order topological insulators and
  phosphorene},}\ }\href {\doibase 10.1103/PhysRevB.98.045125} {\bibfield
  {journal} {\bibinfo  {journal} {Phys. Rev. B}\ }\textbf {\bibinfo {volume}
  {98}},\ \bibinfo {pages} {045125} (\bibinfo {year} {2018})}\BibitemShut
  {NoStop}%
\bibitem [{\citenamefont {Khalaf}\ \emph {et~al.}(2019)\citenamefont {Khalaf},
  \citenamefont {Benalcazar}, \citenamefont {Hughes},\ and\ \citenamefont
  {Queiroz}}]{khalaf-2019arxiv}%
  \BibitemOpen
  \bibfield  {author} {\bibinfo {author} {\bibfnamefont {Eslam}\ \bibnamefont
  {Khalaf}}, \bibinfo {author} {\bibfnamefont {Wladimir~A.}\ \bibnamefont
  {Benalcazar}}, \bibinfo {author} {\bibfnamefont {Taylor~L.}\ \bibnamefont
  {Hughes}}, \ and\ \bibinfo {author} {\bibfnamefont {Raquel}\ \bibnamefont
  {Queiroz}},\ }\href@noop {} {\enquote {\bibinfo {title} {Boundary-obstructed
  topological phases},}\ } (\bibinfo {year} {2019}),\ \Eprint
  {http://arxiv.org/abs/1908.00011} {arXiv:1908.00011 [cond-mat.mes-hall]}
  \BibitemShut {NoStop}%
\bibitem [{\citenamefont {Li}\ \emph {et~al.}(2020)\citenamefont {Li},
  \citenamefont {Zhu}, \citenamefont {Benalcazar},\ and\ \citenamefont
  {Hughes}}]{li-prb20}%
  \BibitemOpen
  \bibfield  {author} {\bibinfo {author} {\bibfnamefont {Tianhe}\ \bibnamefont
  {Li}}, \bibinfo {author} {\bibfnamefont {Penghao}\ \bibnamefont {Zhu}},
  \bibinfo {author} {\bibfnamefont {Wladimir~A.}\ \bibnamefont {Benalcazar}}, \
  and\ \bibinfo {author} {\bibfnamefont {Taylor~L.}\ \bibnamefont {Hughes}},\
  }\bibfield  {title} {\enquote {\bibinfo {title} {Fractional disclination
  charge in two-dimensional ${C}_{n}$-symmetric topological crystalline
  insulators},}\ }\href {\doibase 10.1103/PhysRevB.101.115115} {\bibfield
  {journal} {\bibinfo  {journal} {Phys. Rev. B}\ }\textbf {\bibinfo {volume}
  {101}},\ \bibinfo {pages} {115115} (\bibinfo {year} {2020})}\BibitemShut
  {NoStop}%
\bibitem [{\citenamefont {Schindler}\ \emph {et~al.}(2019)\citenamefont
  {Schindler}, \citenamefont {Brzezi\ifmmode~\acute{n}\else \'{n}\fi{}ska},
  \citenamefont {Benalcazar}, \citenamefont {Iraola}, \citenamefont {Bouhon},
  \citenamefont {Tsirkin}, \citenamefont {Vergniory},\ and\ \citenamefont
  {Neupert}}]{schindler-prresearch19}%
  \BibitemOpen
  \bibfield  {author} {\bibinfo {author} {\bibfnamefont {Frank}\ \bibnamefont
  {Schindler}}, \bibinfo {author} {\bibfnamefont {Marta}\ \bibnamefont
  {Brzezi\ifmmode~\acute{n}\else \'{n}\fi{}ska}}, \bibinfo {author}
  {\bibfnamefont {Wladimir~A.}\ \bibnamefont {Benalcazar}}, \bibinfo {author}
  {\bibfnamefont {Mikel}\ \bibnamefont {Iraola}}, \bibinfo {author}
  {\bibfnamefont {Adrien}\ \bibnamefont {Bouhon}}, \bibinfo {author}
  {\bibfnamefont {Stepan~S.}\ \bibnamefont {Tsirkin}}, \bibinfo {author}
  {\bibfnamefont {Maia~G.}\ \bibnamefont {Vergniory}}, \ and\ \bibinfo {author}
  {\bibfnamefont {Titus}\ \bibnamefont {Neupert}},\ }\bibfield  {title}
  {\enquote {\bibinfo {title} {Fractional corner charges in spin-orbit coupled
  crystals},}\ }\href {\doibase 10.1103/PhysRevResearch.1.033074} {\bibfield
  {journal} {\bibinfo  {journal} {Phys. Rev. Research}\ }\textbf {\bibinfo
  {volume} {1}},\ \bibinfo {pages} {033074} (\bibinfo {year}
  {2019})}\BibitemShut {NoStop}%
\bibitem [{\citenamefont {Watanabe}\ and\ \citenamefont
  {Ono}(2020)}]{watanabe-prb20}%
  \BibitemOpen
  \bibfield  {author} {\bibinfo {author} {\bibfnamefont {Haruki}\ \bibnamefont
  {Watanabe}}\ and\ \bibinfo {author} {\bibfnamefont {Seishiro}\ \bibnamefont
  {Ono}},\ }\bibfield  {title} {\enquote {\bibinfo {title} {Corner charge and
  bulk multipole moment in periodic systems},}\ }\href {\doibase
  10.1103/PhysRevB.102.165120} {\bibfield  {journal} {\bibinfo  {journal}
  {Phys. Rev. B}\ }\textbf {\bibinfo {volume} {102}},\ \bibinfo {pages}
  {165120} (\bibinfo {year} {2020})}\BibitemShut {NoStop}%
\bibitem [{\citenamefont {Kooi}\ \emph {et~al.}(2021)\citenamefont {Kooi},
  \citenamefont {van Miert},\ and\ \citenamefont {Ortix}}]{kooi-npjqm21}%
  \BibitemOpen
  \bibfield  {author} {\bibinfo {author} {\bibfnamefont {Sander}\ \bibnamefont
  {Kooi}}, \bibinfo {author} {\bibfnamefont {Guido}\ \bibnamefont {van Miert}},
  \ and\ \bibinfo {author} {\bibfnamefont {Carmine}\ \bibnamefont {Ortix}},\
  }\bibfield  {title} {\enquote {\bibinfo {title} {{The bulk-corner
  correspondence of time-reversal symmetric insulators}},}\ }\href {\doibase
  10.1038/s41535-020-00300-7} {\bibfield  {journal} {\bibinfo  {journal} {npj
  Quantum Materials}\ }\textbf {\bibinfo {volume} {6}} (\bibinfo {year}
  {2021}),\ 10.1038/s41535-020-00300-7}\BibitemShut {NoStop}%
\bibitem [{\citenamefont {Trifunovic}(2020)}]{trifunovic-prresearch20}%
  \BibitemOpen
  \bibfield  {author} {\bibinfo {author} {\bibfnamefont {Luka}\ \bibnamefont
  {Trifunovic}},\ }\bibfield  {title} {\enquote {\bibinfo {title}
  {Bulk-and-edge to corner correspondence},}\ }\href {\doibase
  10.1103/PhysRevResearch.2.043012} {\bibfield  {journal} {\bibinfo  {journal}
  {Phys. Rev. Research}\ }\textbf {\bibinfo {volume} {2}},\ \bibinfo {pages}
  {043012} (\bibinfo {year} {2020})}\BibitemShut {NoStop}%
\bibitem [{\citenamefont {Watanabe}\ and\ \citenamefont
  {Po}(2020)}]{watanabe-arxiv20}%
  \BibitemOpen
  \bibfield  {author} {\bibinfo {author} {\bibfnamefont {Haruki}\ \bibnamefont
  {Watanabe}}\ and\ \bibinfo {author} {\bibfnamefont {Hoi~Chun}\ \bibnamefont
  {Po}},\ }\href@noop {} {\enquote {\bibinfo {title} {Fractional corner charge
  of sodium chloride},}\ } (\bibinfo {year} {2020}),\ \Eprint
  {http://arxiv.org/abs/2009.04845} {arXiv:2009.04845 [cond-mat.mtrl-sci]}
  \BibitemShut {NoStop}%
\bibitem [{\citenamefont {Kang}\ \emph {et~al.}(2019)\citenamefont {Kang},
  \citenamefont {Shiozaki},\ and\ \citenamefont {Cho}}]{kang-prb19}%
  \BibitemOpen
  \bibfield  {author} {\bibinfo {author} {\bibfnamefont {Byungmin}\
  \bibnamefont {Kang}}, \bibinfo {author} {\bibfnamefont {Ken}\ \bibnamefont
  {Shiozaki}}, \ and\ \bibinfo {author} {\bibfnamefont {Gil~Young}\
  \bibnamefont {Cho}},\ }\bibfield  {title} {\enquote {\bibinfo {title}
  {Many-body order parameters for multipoles in solids},}\ }\href {\doibase
  10.1103/PhysRevB.100.245134} {\bibfield  {journal} {\bibinfo  {journal}
  {Phys. Rev. B}\ }\textbf {\bibinfo {volume} {100}},\ \bibinfo {pages}
  {245134} (\bibinfo {year} {2019})}\BibitemShut {NoStop}%
\bibitem [{\citenamefont {Wheeler}\ \emph {et~al.}(2019)\citenamefont
  {Wheeler}, \citenamefont {Wagner},\ and\ \citenamefont
  {Hughes}}]{wheeler-prb19}%
  \BibitemOpen
  \bibfield  {author} {\bibinfo {author} {\bibfnamefont {William~A.}\
  \bibnamefont {Wheeler}}, \bibinfo {author} {\bibfnamefont {Lucas~K.}\
  \bibnamefont {Wagner}}, \ and\ \bibinfo {author} {\bibfnamefont {Taylor~L.}\
  \bibnamefont {Hughes}},\ }\bibfield  {title} {\enquote {\bibinfo {title}
  {Many-body electric multipole operators in extended systems},}\ }\href
  {\doibase 10.1103/PhysRevB.100.245135} {\bibfield  {journal} {\bibinfo
  {journal} {Phys. Rev. B}\ }\textbf {\bibinfo {volume} {100}},\ \bibinfo
  {pages} {245135} (\bibinfo {year} {2019})}\BibitemShut {NoStop}%
\bibitem [{\citenamefont {Resta}(1998)}]{resta-prl98}%
  \BibitemOpen
  \bibfield  {author} {\bibinfo {author} {\bibfnamefont {R.}~\bibnamefont
  {Resta}},\ }\bibfield  {title} {\enquote {\bibinfo {title}
  {Quantum-mechanical position operator in extended systems},}\ }\href@noop {}
  {\bibfield  {journal} {\bibinfo  {journal} {Phys. Rev. Lett.}\ }\textbf
  {\bibinfo {volume} {80}},\ \bibinfo {pages} {1800--1803} (\bibinfo {year}
  {1998})}\BibitemShut {NoStop}%
\bibitem [{\citenamefont {Ono}\ \emph {et~al.}(2019)\citenamefont {Ono},
  \citenamefont {Trifunovic},\ and\ \citenamefont {Watanabe}}]{ono-prb19}%
  \BibitemOpen
  \bibfield  {author} {\bibinfo {author} {\bibfnamefont {Seishiro}\
  \bibnamefont {Ono}}, \bibinfo {author} {\bibfnamefont {Luka}\ \bibnamefont
  {Trifunovic}}, \ and\ \bibinfo {author} {\bibfnamefont {Haruki}\ \bibnamefont
  {Watanabe}},\ }\bibfield  {title} {\enquote {\bibinfo {title} {Difficulties
  in operator-based formulation of the bulk quadrupole moment},}\ }\href
  {\doibase 10.1103/PhysRevB.100.245133} {\bibfield  {journal} {\bibinfo
  {journal} {Phys. Rev. B}\ }\textbf {\bibinfo {volume} {100}},\ \bibinfo
  {pages} {245133} (\bibinfo {year} {2019})}\BibitemShut {NoStop}%
\bibitem [{\citenamefont {Marzari}\ and\ \citenamefont
  {Vanderbilt}(1997)}]{marzari-prb97}%
  \BibitemOpen
  \bibfield  {author} {\bibinfo {author} {\bibfnamefont {N.}~\bibnamefont
  {Marzari}}\ and\ \bibinfo {author} {\bibfnamefont {D.}~\bibnamefont
  {Vanderbilt}},\ }\bibfield  {title} {\enquote {\bibinfo {title} {{Maximally
  localized generalized Wannier functions for composite energy bands}},}\
  }\href {\doibase 10.1103/PhysRevB.56.12847} {\bibfield  {journal} {\bibinfo
  {journal} {Phys. Rev. B}\ }\textbf {\bibinfo {volume} {56}},\ \bibinfo
  {pages} {12847} (\bibinfo {year} {1997})}\BibitemShut {NoStop}%
\bibitem [{\citenamefont {Marzari}\ \emph {et~al.}(2012)\citenamefont
  {Marzari}, \citenamefont {Mostofi}, \citenamefont {Yates}, \citenamefont
  {Souza},\ and\ \citenamefont {Vanderbilt}}]{marzari-rmp12}%
  \BibitemOpen
  \bibfield  {author} {\bibinfo {author} {\bibfnamefont {Nicola}\ \bibnamefont
  {Marzari}}, \bibinfo {author} {\bibfnamefont {Arash~A.}\ \bibnamefont
  {Mostofi}}, \bibinfo {author} {\bibfnamefont {Jonathan~R.}\ \bibnamefont
  {Yates}}, \bibinfo {author} {\bibfnamefont {Ivo}\ \bibnamefont {Souza}}, \
  and\ \bibinfo {author} {\bibfnamefont {David}\ \bibnamefont {Vanderbilt}},\
  }\bibfield  {title} {\enquote {\bibinfo {title} {Maximally localized
  {Wannier} functions: {Theory} and applications},}\ }\href {\doibase
  10.1103/RevModPhys.84.1419} {\bibfield  {journal} {\bibinfo  {journal} {Rev.
  Mod. Phys.}\ }\textbf {\bibinfo {volume} {84}},\ \bibinfo {pages}
  {1419--1475} (\bibinfo {year} {2012})}\BibitemShut {NoStop}%
\bibitem [{\citenamefont {Zhou}\ \emph {et~al.}(2015)\citenamefont {Zhou},
  \citenamefont {Rabe},\ and\ \citenamefont {Vanderbilt}}]{zhou-prb15}%
  \BibitemOpen
  \bibfield  {author} {\bibinfo {author} {\bibfnamefont {Yuanjun}\ \bibnamefont
  {Zhou}}, \bibinfo {author} {\bibfnamefont {Karin~M.}\ \bibnamefont {Rabe}}, \
  and\ \bibinfo {author} {\bibfnamefont {David}\ \bibnamefont {Vanderbilt}},\
  }\bibfield  {title} {\enquote {\bibinfo {title} {Surface polarization and
  edge charges},}\ }\href {\doibase 10.1103/PhysRevB.92.041102} {\bibfield
  {journal} {\bibinfo  {journal} {Phys. Rev. B}\ }\textbf {\bibinfo {volume}
  {92}},\ \bibinfo {pages} {041102} (\bibinfo {year} {2015})}\BibitemShut
  {NoStop}%
\bibitem [{\citenamefont {Vanderbilt}\ and\ \citenamefont
  {King-Smith}(1993)}]{vanderbilt-prb93}%
  \BibitemOpen
  \bibfield  {author} {\bibinfo {author} {\bibfnamefont {David}\ \bibnamefont
  {Vanderbilt}}\ and\ \bibinfo {author} {\bibfnamefont {R.~D.}\ \bibnamefont
  {King-Smith}},\ }\bibfield  {title} {\enquote {\bibinfo {title} {Electric
  polarization as a bulk quantity and its relation to surface charge},}\
  }\href@noop {} {\bibfield  {journal} {\bibinfo  {journal} {Phys. Rev. B}\
  }\textbf {\bibinfo {volume} {48}},\ \bibinfo {pages} {4442--4455} (\bibinfo
  {year} {1993})}\BibitemShut {NoStop}%
\bibitem [{\citenamefont {Trifunovic}\ \emph {et~al.}(2019)\citenamefont
  {Trifunovic}, \citenamefont {Ono},\ and\ \citenamefont
  {Watanabe}}]{trifunovic-prb19}%
  \BibitemOpen
  \bibfield  {author} {\bibinfo {author} {\bibfnamefont {Luka}\ \bibnamefont
  {Trifunovic}}, \bibinfo {author} {\bibfnamefont {Seishiro}\ \bibnamefont
  {Ono}}, \ and\ \bibinfo {author} {\bibfnamefont {Haruki}\ \bibnamefont
  {Watanabe}},\ }\bibfield  {title} {\enquote {\bibinfo {title} {Geometric
  orbital magnetization in adiabatic processes},}\ }\href {\doibase
  10.1103/PhysRevB.100.054408} {\bibfield  {journal} {\bibinfo  {journal}
  {Phys. Rev. B}\ }\textbf {\bibinfo {volume} {100}},\ \bibinfo {pages}
  {054408} (\bibinfo {year} {2019})}\BibitemShut {NoStop}%
\bibitem [{\citenamefont {Daido}\ \emph {et~al.}(2020)\citenamefont {Daido},
  \citenamefont {Shitade},\ and\ \citenamefont {Yanase}}]{daido-prb20}%
  \BibitemOpen
  \bibfield  {author} {\bibinfo {author} {\bibfnamefont {Akito}\ \bibnamefont
  {Daido}}, \bibinfo {author} {\bibfnamefont {Atsuo}\ \bibnamefont {Shitade}},
  \ and\ \bibinfo {author} {\bibfnamefont {Youichi}\ \bibnamefont {Yanase}},\
  }\bibfield  {title} {\enquote {\bibinfo {title} {Thermodynamic approach to
  electric quadrupole moments},}\ }\href {\doibase 10.1103/PhysRevB.102.235149}
  {\bibfield  {journal} {\bibinfo  {journal} {Phys. Rev. B}\ }\textbf {\bibinfo
  {volume} {102}},\ \bibinfo {pages} {235149} (\bibinfo {year}
  {2020})}\BibitemShut {NoStop}%
\bibitem [{pyt()}]{pythtb}%
  \BibitemOpen
  \href@noop {} {}\bibinfo {note} {The \textsc{PythTB} code package is
  available at http://www.physics.rutgers.edu/pythtb/about.html}\BibitemShut
  {NoStop}%
\bibitem [{\citenamefont {Resta}(2010)}]{resta-prl10}%
  \BibitemOpen
  \bibfield  {author} {\bibinfo {author} {\bibfnamefont {Raffaele}\
  \bibnamefont {Resta}},\ }\bibfield  {title} {\enquote {\bibinfo {title}
  {Towards a bulk theory of flexoelectricity},}\ }\href {\doibase
  10.1103/PhysRevLett.105.127601} {\bibfield  {journal} {\bibinfo  {journal}
  {Phys. Rev. Lett.}\ }\textbf {\bibinfo {volume} {105}},\ \bibinfo {pages}
  {127601} (\bibinfo {year} {2010})}\BibitemShut {NoStop}%
\bibitem [{\citenamefont {Liu}\ and\ \citenamefont
  {Vanderbilt}(2014)}]{liu-prb14}%
  \BibitemOpen
  \bibfield  {author} {\bibinfo {author} {\bibfnamefont {Jianpeng}\
  \bibnamefont {Liu}}\ and\ \bibinfo {author} {\bibfnamefont {David}\
  \bibnamefont {Vanderbilt}},\ }\bibfield  {title} {\enquote {\bibinfo {title}
  {Spin-orbit spillage as a measure of band inversion in insulators},}\ }\href
  {\doibase 10.1103/PhysRevB.90.125133} {\bibfield  {journal} {\bibinfo
  {journal} {Phys. Rev. B}\ }\textbf {\bibinfo {volume} {90}},\ \bibinfo
  {pages} {125133} (\bibinfo {year} {2014})}\BibitemShut {NoStop}%
\bibitem [{\citenamefont {Benalcazar}\ \emph {et~al.}(2019)\citenamefont
  {Benalcazar}, \citenamefont {Li},\ and\ \citenamefont
  {Hughes}}]{benalcazar-prb19}%
  \BibitemOpen
  \bibfield  {author} {\bibinfo {author} {\bibfnamefont {Wladimir~A.}\
  \bibnamefont {Benalcazar}}, \bibinfo {author} {\bibfnamefont {Tianhe}\
  \bibnamefont {Li}}, \ and\ \bibinfo {author} {\bibfnamefont {Taylor~L.}\
  \bibnamefont {Hughes}},\ }\bibfield  {title} {\enquote {\bibinfo {title}
  {Quantization of fractional corner charge in ${C}_{n}$-symmetric higher-order
  topological crystalline insulators},}\ }\href {\doibase
  10.1103/PhysRevB.99.245151} {\bibfield  {journal} {\bibinfo  {journal} {Phys.
  Rev. B}\ }\textbf {\bibinfo {volume} {99}},\ \bibinfo {pages} {245151}
  (\bibinfo {year} {2019})}\BibitemShut {NoStop}%
\bibitem [{\citenamefont {Thonhauser}\ \emph {et~al.}(2005)\citenamefont
  {Thonhauser}, \citenamefont {Ceresoli}, \citenamefont {Vanderbilt},\ and\
  \citenamefont {Resta}}]{thonhauser-prl05}%
  \BibitemOpen
  \bibfield  {author} {\bibinfo {author} {\bibfnamefont {T.}~\bibnamefont
  {Thonhauser}}, \bibinfo {author} {\bibfnamefont {Davide}\ \bibnamefont
  {Ceresoli}}, \bibinfo {author} {\bibfnamefont {David}\ \bibnamefont
  {Vanderbilt}}, \ and\ \bibinfo {author} {\bibfnamefont {R.}~\bibnamefont
  {Resta}},\ }\bibfield  {title} {\enquote {\bibinfo {title} {Orbital
  magnetization in periodic insulators},}\ }\href {\doibase
  {10.1103/PhysRevLett.95.137205}} {\bibfield  {journal} {\bibinfo  {journal}
  {Phys. Rev. Lett}\ }\textbf {\bibinfo {volume} {95}},\ \bibinfo {pages}
  {137205} (\bibinfo {year} {2005})}\BibitemShut {NoStop}%
\bibitem [{\citenamefont {Ceresoli}\ \emph {et~al.}(2006)\citenamefont
  {Ceresoli}, \citenamefont {Thonhauser}, \citenamefont {Vanderbilt},\ and\
  \citenamefont {Resta}}]{ceresoli-prb06}%
  \BibitemOpen
  \bibfield  {author} {\bibinfo {author} {\bibfnamefont {D.}~\bibnamefont
  {Ceresoli}}, \bibinfo {author} {\bibfnamefont {T.}~\bibnamefont
  {Thonhauser}}, \bibinfo {author} {\bibfnamefont {D.}~\bibnamefont
  {Vanderbilt}}, \ and\ \bibinfo {author} {\bibfnamefont {R.}~\bibnamefont
  {Resta}},\ }\bibfield  {title} {\enquote {\bibinfo {title} {Orbital
  magnetization in crystalline solids: {Multi-band} insulators, {Chern}
  insulators, and metals},}\ }\href {\doibase {10.1103/PhysRevB.74.024408}}
  {\bibfield  {journal} {\bibinfo  {journal} {Phys. Rev. B}\ }\textbf {\bibinfo
  {volume} {74}},\ \bibinfo {pages} {024408} (\bibinfo {year}
  {2006})}\BibitemShut {NoStop}%
\end{thebibliography}%

\end{document}